\documentclass[a4paper,11pt]{article}
\pdfoutput=1 % if your are submitting a pdflatex (i.e. if you have
\usepackage{jheppub} % for details on the use of the package, please
\usepackage[T1]{fontenc} % if needed

\input{defs}
\newcommand{\figref}[1]       {figure~\ref{fig:#1}}
\newcommand{\Figref}[1]       {Figure~\ref{fig:#1}}
\newcommand{\figsref}[2]      {figures~\ref{fig:#1} and \ref{fig:#2}}
\newcommand{\Figsref}[2]      {Figures~\ref{fig:#1} and \ref{fig:#2}}

\newcommand{\Figrange}[2]     {Figures~\ref{fig:#1}--\ref{fig:#2}}
\newcommand{\eref}[1]         {eq.~\ref{eq:#1}}

\newcommand{\secref}[1]       {section~\ref{sec:#1}}

\newcommand{\appref}[1]       {appendix~\ref{app:#1}}

\newcommand{\bibref}[1]       {ref.~\cite{#1}}

\usepackage{xspace}
\usepackage{graphicx}
\usepackage{subcaption}
\usepackage{relsize}

\title{\boldmath Pileup density estimate independent on jet multiplicity}

\author[a,1]{P. Berta,\note{Corresponding author.}}
\author[b]{J. Smie\v sko,}
\author[a]{M. Spousta}

\affiliation[a]{Institute of Particle and Nuclear Physics, Faculty of Mathematics and Physics, Charles University, V Hole\v sovi\v ck\' ach 2, 180 00 Prague 8, Czech Republic}
\affiliation[b]{CERN, 1211 Geneva 23, Switzerland}

\emailAdd{peter.berta@matfyz.cuni.cz}
\emailAdd{juraj.smiesko@cern.ch}
\emailAdd{Martin.Spousta@matfyz.cuni.cz}

\abstract{

The hard-scatter processes in hadronic collisions are often largely contaminated with soft background coming from pileup in proton-proton collisions, or underlying event in heavy-ion collisions. There are multiple methods to remove the effect of pileup for jets. Two such methods, Area Subtraction and Constituent Subtraction, use the pileup density as the main ingredient to estimate the magnitude of pileup contribution on an event-by-event basis. The state-of-the-art approaches to estimating pileup density are sensitive to the number of hard-scatter jets in the event. This paper presents a new pileup-density estimation method that minimizes the sensitivity on the presence of hard-scatter jets in the event. Using a detector-level simulation, we provide a comparison of the new method with the state-of-the-art estimation methods. We observe a significantly lower bias for the estimated pileup density when using the new method. We conclude that the new method has the potential to significantly improve pileup mitigation in proton-proton collisions or the underlying event subtraction in heavy-ion collisions.

}

\begin{document} 

\maketitle
\flushbottom

\section{Introduction}
\label{sec:intro}
Increasing beam intensities at the LHC led to the increasing number of simultaneous proton-proton collisions occurring within one crossing of beam bunches, also known as pileup. The average number of simultaneous collisions increased from ${\sim}15$, in the early stage of LHC run 2, up to ${\sim}35$, in the last stage \cite{ref:pileup1,ref:pileup2}.
The pileup of 35 collisions contributes on average ${\sim}1.5$~GeV to transverse momentum ($p_T$) of a typical jet with radius 0.4 reconstructed in
pseudorapidity-azimuth ($\eta-\phi$) space. Pileup, therefore, influences all measurements involving jets and reconstruction of missing transverse energy and may lead to significant biases if no correction is applied. Correcting for the pileup using pileup mitigation techniques is thus a crucial task for the LHC physics program \cite{Soyez:2018opl}.   
The mitigation of pileup is a non-simple problem due to a broad distribution of the number of pileup collisions, fluctuating and $\eta$-dependent $p_T$ originating in pileup collisions, and a simultaneous presence of hard processes.

Pileup mitigation techniques may be divided into two classes. 
The first class consists of techniques that do not alter the definition of the jet. In the most straightforward case, these techniques apply an additive or multiplicative
correction on reconstructed jet $\pt$ or four-momentum. The representative of these is the \areaSubtraction method~\cite{Cacciari:2007fd,Cacciari:2008gn} and its extensions used in experiments \cite{ATLAS:2015ull,CMS:2012rmf,CMS:2016lmd,ATLAS:2020cli}. Since \areaSubtraction method does not allow correcting the jet substructure, methods that correct the pileup contribution at the level of jet constituents were developed. Widely used representatives of these are (Iterative) Constituent Subtraction method (ICS or CS) \cite{Berta:2014eza,Berta:2019hnj}, SoftKiller~\cite{Cacciari:2014gra}, PUPPI~\cite{Bertolini:2014bba,CMS:2020ebo}, and jet cleansing~\cite{Krohn:2013lba}.
An alternative to these is the Shape Expansion method \cite{Soyez:2012hv}, which allows correcting individual substructure-based observables. The second class of mitigation techniques is the so-called groomers. All of these methods recluster the constituents of jets and apply so-called filtering \cite{Butterworth:2008iy}, trimming \cite{Krohn:2009th}, pruning \cite{Ellis:2009me}, or 
soft drop technique \cite{Larkoski:2014wba}
to isolate certain jet topologies and mitigate the impact of soft backgrounds. 

Many pileup techniques use as an input the estimate of 
pileup $\pt$ density, $\rho$ (an exception is e.g. PUPPI). 
Commonly used techniques to determine $\rho$ are discussed in section~\ref{sec:stateOfTheArt}. While these techniques are used by all four major LHC experiments, they exhibit residual dependencies on internal parameters as well as on $n_{PU}$ 
\cite{CMS:2016lmd,ATLAS:2020cli,Soyez:2018opl}. The residual $n_{PU}$ dependence is dealt with in each LHC experiment differently, but in general, the $\rho$ estimate can be corrected by an $n_{PU}$-dependent correction factor. 
A complication to this approach of correcting $\rho$ is brought by the fact that $\rho$ further depends on the presence of a particular hard-scattering process, as discussed e.g. in \cite{Soyez:2018opl}. This is connected with jets being included as a part of the background or with different amounts of parton radiation leaking out of the jet cone in different hard-scattering processes. Consequently, the $n_{PU}$-dependence of $\rho$ is process-dependent as further quantified in section~\ref{sec:current_performance}. Since it is not possible to derive pileup correction separately for each measurement involving a particular hard-scattering process, it is highly desirable to implement a commonly usable method that would allow reducing significantly the process-dependence of $\rho$.
This is the main objective of the new method for the $\rho$ estimation presented in this paper. This new method is described in section~\ref{sec:newRho}, it is named \mname, and it is implemented in \texttt{FastJet Contrib} \cite{Salam:2019:fastjetContrib} package in version 1.054 or higher. The performance of the new method is quantified in section~\ref{sec:performance}.

A problem similar to pileup is also present in ultra-relativistic collisions of heavy ions performed at the LHC and RHIC colliders due to the underlying event. While we concentrate solely on the pileup in $pp$ collisions in this paper, the method presented here may also be used in the reconstruction of jets in heavy-ion collisions, where it might reduce the impact of jets on the estimate of $\rho$. Studying this goes, however, beyond the scope of this paper.

\section{Test samples and configuration of subtraction}
\label{sec:setup}
The performance of subtraction methods is evaluated in simulated $pp$ collisions with center-of-mass energy $\sqrt{s}=14\tev$. Four samples are used in this study to quantify the jet multiplicity dependence of the pileup subtraction. Samples that use PYTHIA use PYTHIA~8.303~\cite{Sjostrand:2006za,Sjostrand:2007gs}, tune 4C, and CTEQ~5L parton density functions (PDFs)~\cite{Lai:1999wy}. Sample that uses MadGraph uses MadGraph5\_\-aMC@NLO 3.3.1~\cite{Alwall:2014hca} and the same PYTHIA configuration as other samples to generate parton shower, hadronization, and hadron decays. Samples with hard-scatter process (signal) are generated without the underlying event. The four samples are following: 
\begin{itemize}
\item Four-top-quark (\tttt) sample: The hard-scatter process is generated using MadGraph with subsequent running of PYTHIA. Only hadronic decays of W bosons are allowed.
\item Top-antitop (\ttbar) sample: Generated with PYTHIA. Only hadronic decays of W bosons are allowed.
\item Dijet sample: Generated with PYTHIA. The minimum transverse momentum of partons outgoing the binary process, $\widehat{p}_T^{~\mathrm{min}}$, is set to $30\gev$.
\item No-hard-scatter sample: No hard-scatter process is generated. The \pileup events are overlaid as described below. The signal is not defined in this sample.
\end{itemize}

The events from above samples are referred to as \textit{signal} events. To obtain the \textit{signal+\pileup} events with \pileup included, the signal events are overlaid with inclusive $pp$ collision events. The inclusive $pp$ collision events are generated with~PYTHIA configured to enable all soft QCD processes. The number of overlaid inclusive events, \npu, has a uniform distribution from 0 to 220. Each of four signal+\pileup samples consists of 100 000 events.

A pseudo-detector simulation is used for both signal and signal+\pileup samples. All particles are grouped into towers of size $0.1 \times 0.1$ in the pseudorapidity-azimuth (\etaphi) space. The tower energy is obtained as the sum of the energies of particles pointing to that tower. All neutrinos and muons are discarded. Only towers with $|\eta|<4.0$ are selected. The mass of each tower is set to zero. The $\eta$ and $\phi$ of each tower are randomly smeared using a Gaussian kernel with a width of 0.03 (but maximally up to 0.05).

Besides directly comparing the estimates of $\rho$, we also evaluate the impact of determining $\rho$ on jet energy scale and resolution after \pileup correction. To do that, the signal+\pileup events are corrected with the Jet-by-jet Constituent Subtraction method using various estimated $\rho$. The Jet-by-jet CS uses parameters $\DeltaRmax=\infty$, $\alpha=0$, and $\Aghost=0.0025$. To evaluate the performance of the various methods, the jets from the corrected events are compared to jets from the signal events using the quantities defined in \secref{jet}. All jets are clustered with the \akt algorithm \cite{Cacciari:2008gp} with the distance parameter of $0.4$. Only jets with $|\eta|<3$ and $\pt>20\gev$ are used. All jet finding and background estimation is performed using \texttt{FastJet} 3.4.2~\cite{Cacciari:2011ma, Cacciari:2005hq}.

\section{Pileup density estimation - state of the art}
\label{sec:stateOfTheArt}
The basic ingredient of \areaSubtraction or Constituent-Subtraction-based methods is the background \pt density, $\rho$, which was introduced within the \areaSubtraction method \cite{Cacciari:2007fd}. Several methods to estimate this quantity are described in \cite{Cacciari:2011ma}. In general, $\rho$ can be estimated as a function of other variables, most commonly as a function of rapidity. The estimated $\rho$ is then used to scale the \pt of the \textit{ghosts} in a given \pileup correction method (\areaSubtraction, Shape-expansion, Jet-by-jet \cs, Event-wide \cs, or \ics).

The most common method to estimate $\rho$ is the following:
\begin{enumerate}
\item Break the event into patches of a similar area. Typically, one can break the event into a grid of cells in the \yphi plane (done within the so-called \textit{grid–median} estimator). Alternatively, one can use \kt or \ca jets obtained with a jet radius potentially different from the one used to cluster the physical jets (done within \textit{jet–median} estimator).
\item Calculate the momentum, \ptpatchi, and area, \areapatchi, for each patch $i$. If patches are grid cells, one can use the scalar \pt sum of all particles in patch $i$ and the geometric area in \yphi plane.
\item The estimate of $\rho$ is given by the median of \ptpatchi over \areapatchi ratio determined from a set of all patches:
\begin{equation}
\rhoest = \underset{{i\in \mathrm{patches}}}{\mathrm{median}} \left(\ptpatchi/\areapatchi\right)
\label{rho_scalar}
\end{equation}
\end{enumerate}

The grid-median method has one free parameter called \textit{grid size}. It is the size of the grid cell and its recommended value from \bibref{Soyez:2018opl} is between 0.5 and 0.7 with the optimal recommended value of 0.55. The jet-median method was used in multiple LHC publications where it used \kt jets with the distance parameter of 0.4 (see e.g. \cite{ATLAS:2023tyv,ALICE:2023ama}). Both methods are implemented in the FastJet package \cite{Cacciari:2011ma}.

Optionally, a rescaling can be used to determine $\rho$ based on an expected dependence of $\rho$ in the \yphi plane. Typically, the rescaling is done to take into account the known rapidity dependence of $\rho$, $r(y)$, as described in \bibref{Soyez:2018opl}. When rapidity rescaling is used, each particle \pt in Step 2 is divided by factor $r(y)$, and then the final estimated $\rho$ for rapidity $y$ is evaluated as:
\begin{equation}
\rhoest (y) = r(y) \cdot \underset{{i\in \mathrm{patches}}}{\mathrm{median}} \left(\ptpatchi/\areapatchi\right).
\label{eq:rho_rapDep}
\end{equation}
The rescaling function $r(y)$ can be represented by one-dimensional histogram binned in $y$ where each particle from a sample of $N$ \pileup events is filled in a given $y$ bin with a $\pt$ weight ($N$ large, here
$N \geq 10^5$). This histogram is then scaled by $1/(2\pi \Delta y N)$ where $\Delta y$ is the bin width\footnote{The overall normalization of $r(y)$ has no impact on $\rhoest (y)$. The chosen normalization ensures that $r(y)$ captures the average rapidity dependence of $\rho$ for one \pileup event. However, this normalization also implies that $\ptpatchi$ is unitless in equation \eqref{eq:rho_rapDep} (area in $y \times \phi$ space is considered unitless).}. 

%------------------------------ 
\begin{figure} [h]
  \centering
  \begin{subfigure}{0.48\textwidth}
    \includegraphics[width=\textwidth]{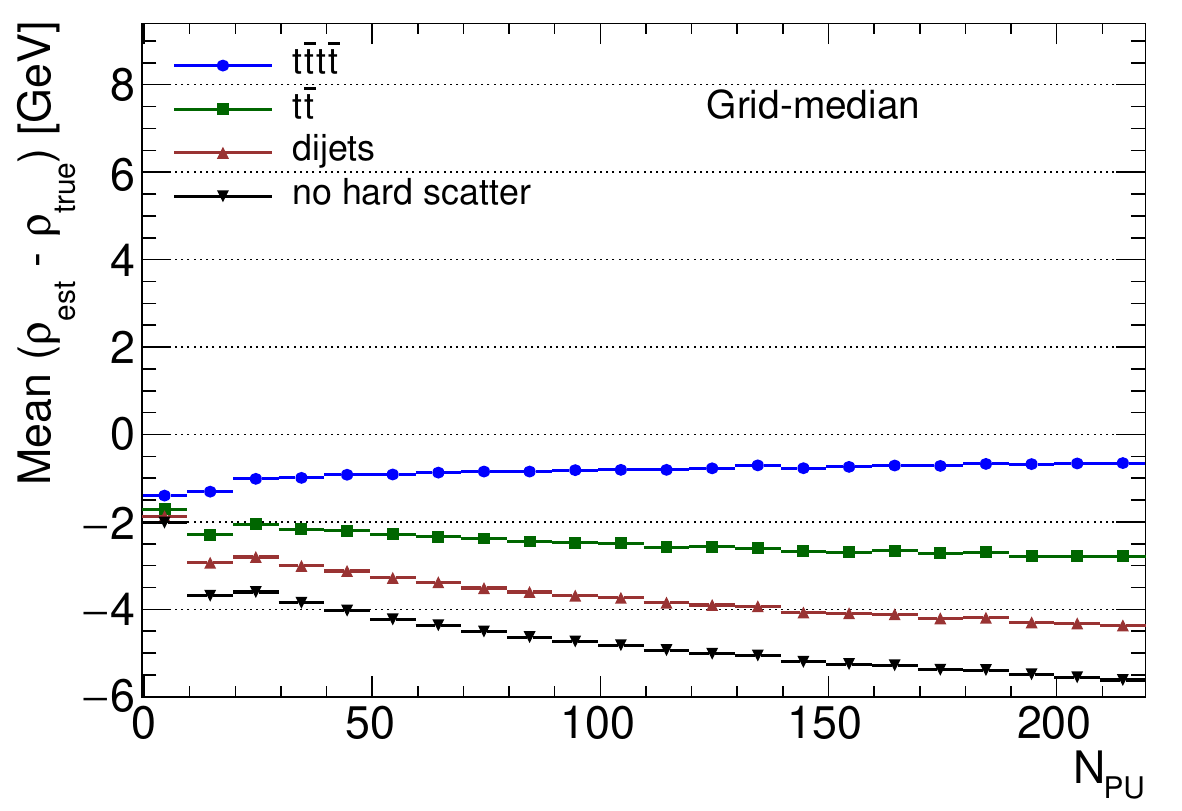}
    \caption{grid-median size 0.3}
    \label{fig:grid:0.3}
  \end{subfigure}
  \begin{subfigure}{0.48\textwidth}
    \includegraphics[width=\textwidth]{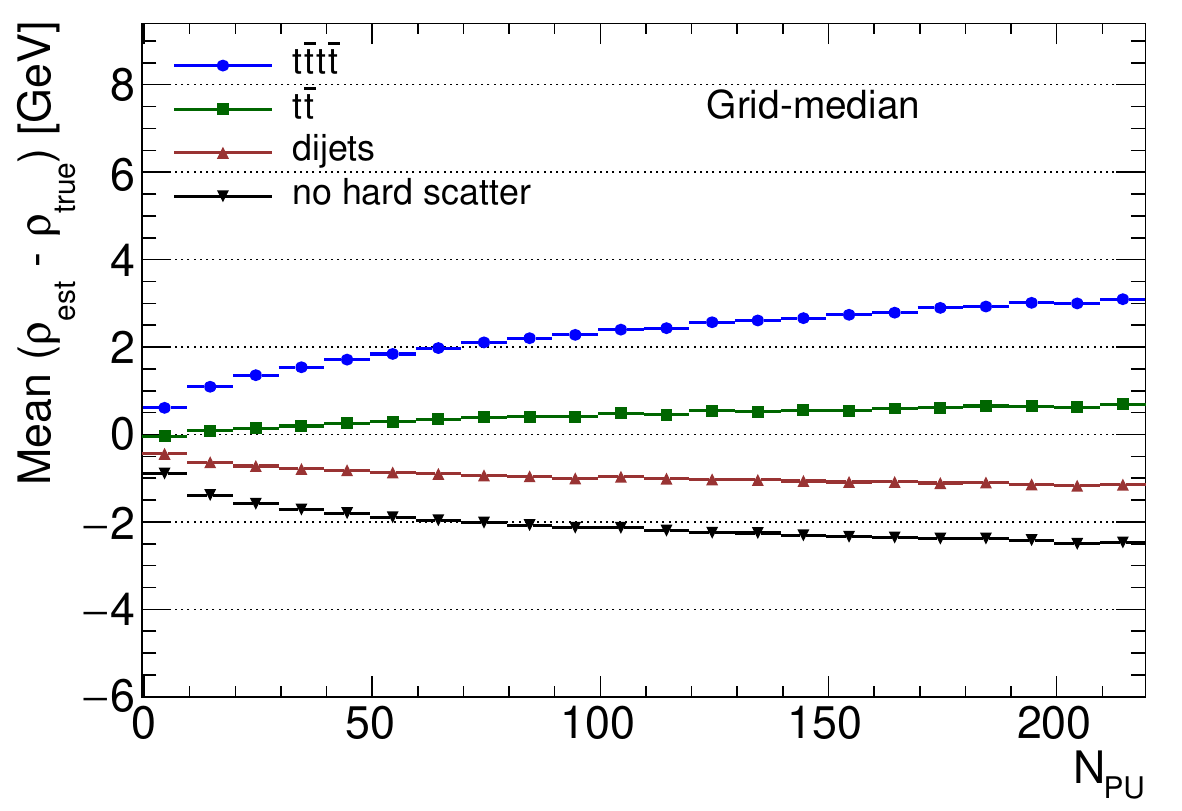}
    \caption{grid-median size 0.55}
    \label{fig:grid:0.55}
  \end{subfigure}
  \begin{subfigure}{0.48\textwidth}
    \includegraphics[width=\textwidth]{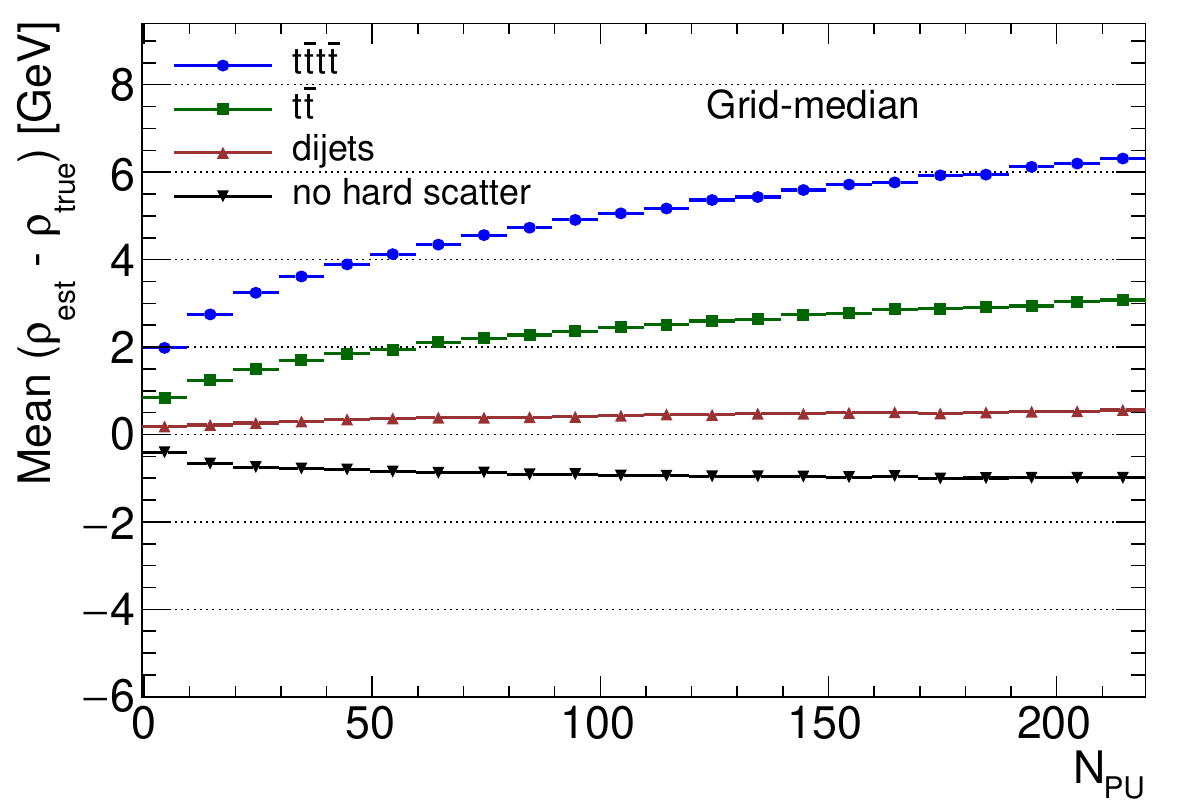}
    \caption{grid-median size 1.0}
    \label{fig:grid:1.0}
  \end{subfigure}
  \begin{subfigure}{0.48\textwidth}
    \includegraphics[width=\textwidth]{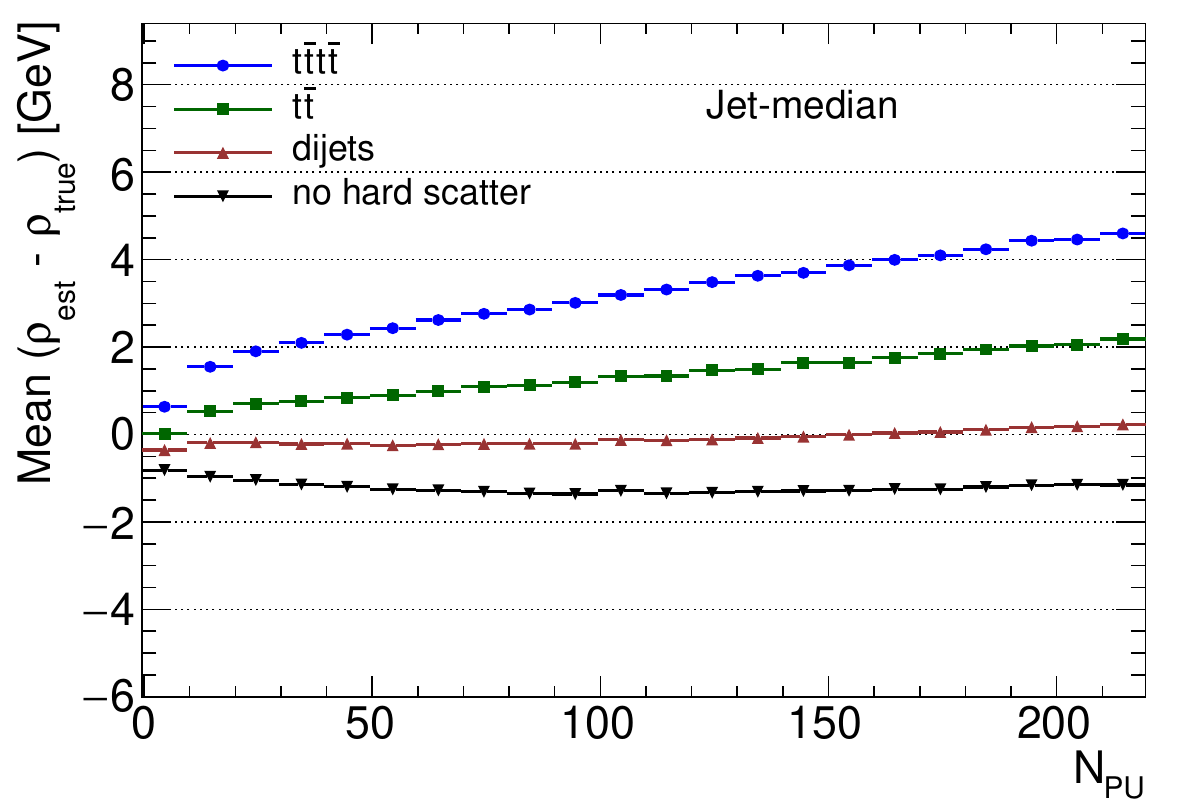}
    \caption{jet-median using \kt $R=0.4$}
    \label{fig:jet:0.4}
  \end{subfigure}
  \caption{Dependence of $\rho$ bias on \npu for $\rho$ obtained using the grid-median method (for three grid size values: 0.3 (a), 0.55 (b), and 1.0 (c)) and using the jet-median method with \kt $R=0.4$ jets (d). The distributions are shown for four samples: no hard-scatter, dijet, \ttbar, and four top quark sample.}  
  \label{fig:gridStandard}
\end{figure}
%%------------------------------ 

\subsection{Performance of the current methods}
\label{sec:current_performance}
An unbiased $\rho$ estimate, \rhotrue, is defined to evaluate how precisely the $\rho$ estimation methods work. The \rhotrue is obtained using purely particles from the pileup. In each event, these pileup particles undergo the pseudo-detector simulation described in \secref{setup}, and the obtained towers are used to determine \rhotrue as a function of rapidity $y$:
\begin{equation}
\rhotrue (y) = r(y) \cdot \left( \sum_i \pti / r(y_i) \right) / A,
\label{eq:rapidity_rescaling}
\end{equation}
where the sum runs over all inputs (towers), \pti and $y_i$ are the transverse momentum and rapidity of particle $i$, respectively. The area $A$ is the total area where the input particles are distributed ($2\pi\cdot2\cdot4\approx 50.3$ in our case) and $r(y)$ is defined above.

The grid-median estimate of $\rho$ is calculated as a function of $y$ using the \texttt{GridMedian\-BackgroundEstimator} tool from \texttt{FastJet} using rapidity rescaling. The jet-median estimate of $\rho$ is calculated as a function of $y$ using the \texttt{JetMedianBackgroundEstimator} tool from \texttt{FastJet} using \kt clustering algorithm with $R=0.4$ to cluster patches. %All values of $\rho$ are evaluated for rapidity $y=0$.

The estimated $\rho$ depends on the grid size and the hard-scatter process as it was discussed in \bibref{Soyez:2018opl}. An extended study with more samples and extended range for \npu is provided in \figref{gridStandard} where the bias on $\rho$ (evaluated at rapidity $y = 0$) is investigated as a function of \npu for four samples and three grid sizes. The grid size has a large impact on the estimated $\rho$. With decreasing grid sizes, the estimated $\rho$ gets increasingly underestimated even in the case of the absence of a hard-scatter process. With increasing grid size, the estimated $\rho$ is less underestimated and it converges to the true value $\rhotrue$ for the sample without a hard-scatter process. This behavior was also observed in \bibref{Soyez:2018opl} where it was explained based on a simple analytic \pileup model. The presence of the hard-scatter process then increases the estimated $\rho$.
With increasing jet multiplicity in the hard-scatter event, the $\rho$ estimate increases compared to the $\rho$ estimate obtained in a sample with no hard-scatter process. In general, the differences in $\rho$ are significant. The difference of 4\gev in the estimated $\rho$ between the \tttt sample and sample without hard-scatter process is seen even for $\npu=35$. These differences then propagate also to the performance of the relevant \pileup correction method.

An example of the impact of this bias on measured quantities can be represented by inclusive jet cross-section. For $R=0.4$ jets in mid-rapidity at the LHC energies, an overestimation of jet \pt by 2 GeV, which is typical for $R=0.4$ jets reconstructed in events with $\npu = 100$, leads to a depletion in the jet cross-section by 10\%.

\begin{figure}[!h]
  \centering
  \begin{subfigure}{0.48\textwidth}
    \includegraphics[width=\textwidth]{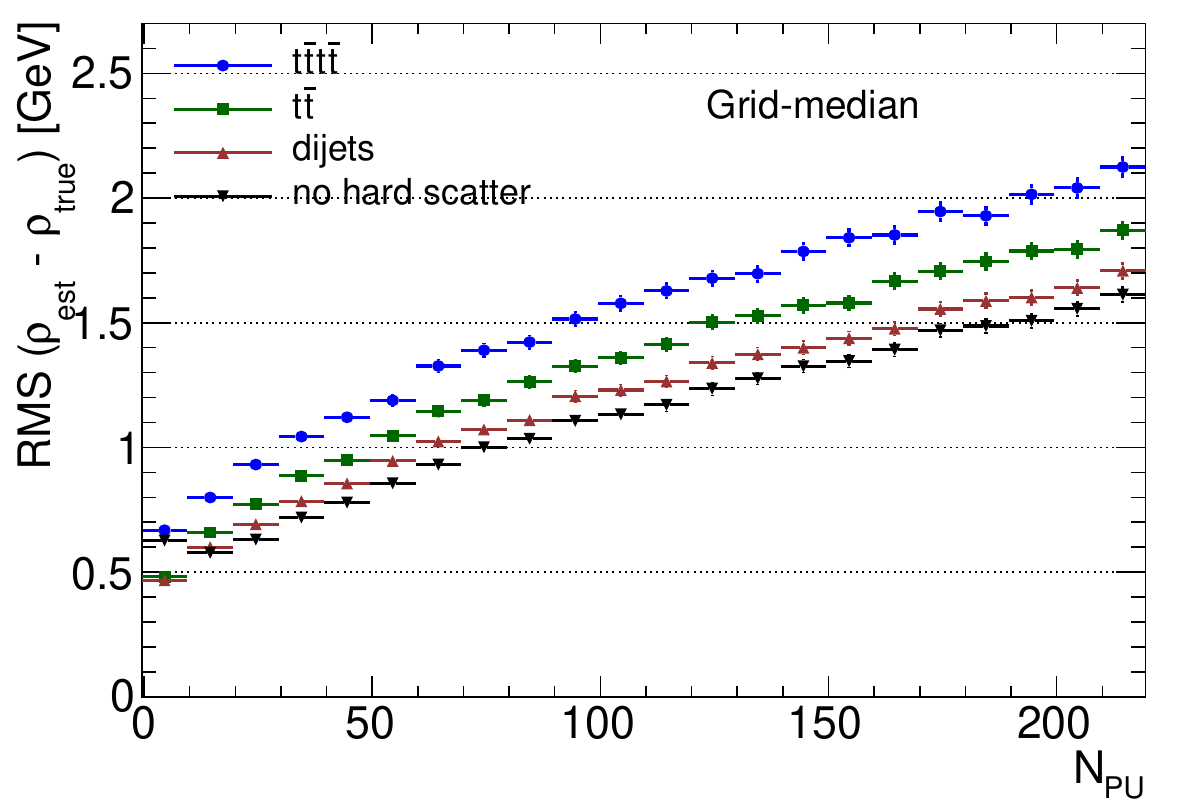}
    \caption{grid-median size 0.55}
    \label{fig:rms_grid:0.55}
  \end{subfigure}
  \begin{subfigure}{0.48\textwidth}
    \includegraphics[width=\textwidth]{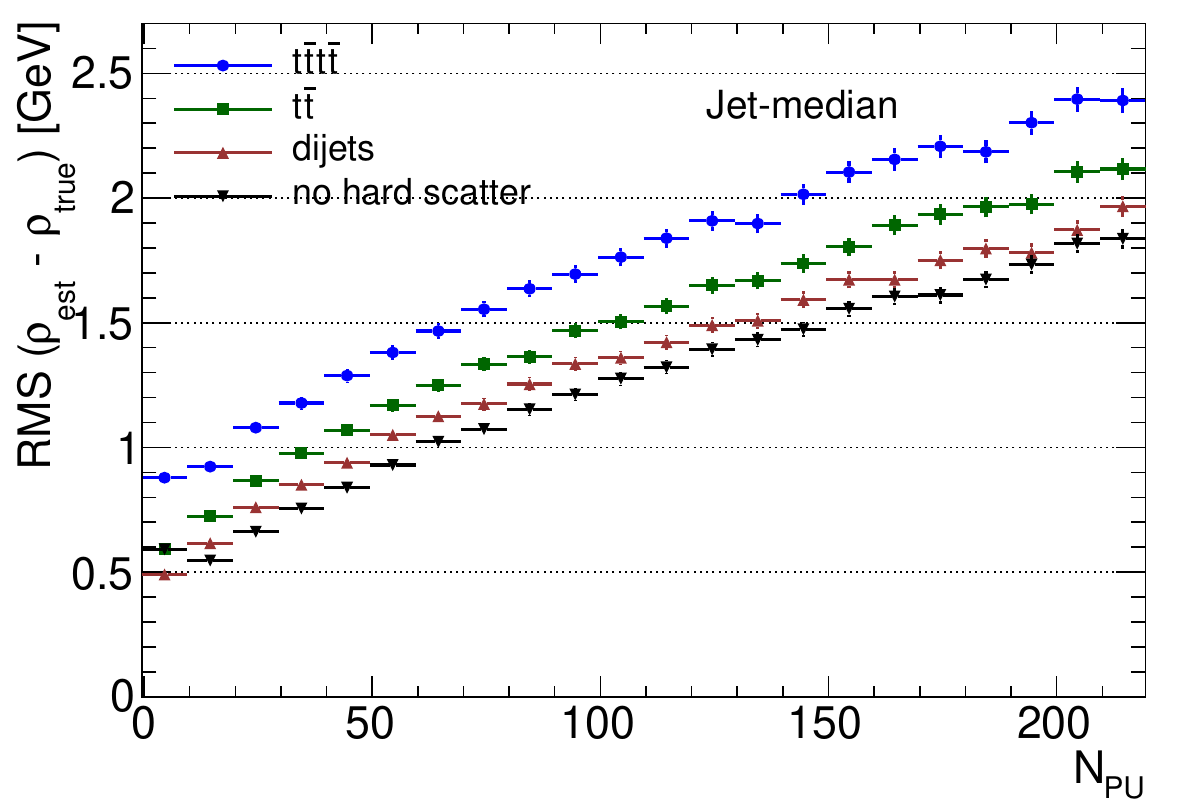}
    \caption{jet-median using \kt R=0.4}
    \label{fig:rms_jet:0.4}
  \end{subfigure}
  \caption{Dependence of $\rho$ resolution on \npu for $\rho$ obtained using the grid-median method with a grid size of 0.55 (a) and jet-median method with \kt $R=0.4$ jets (d). The distributions are shown for four samples: no hard-scatter, dijet, \ttbar, and four top quark sample.}  
  \label{fig:rmsGridStandard}
\end{figure}

Not only the mean difference between the $\rho$ estimates depends on hard-scatter process and \npu but also the resolution of $\rho$ exhibits this
dependence. This is quantified in \figref{rmsGridStandard}. One can see that the resolution differs by almost 50\% at low $\npu$. The worst resolution is achieved for the \tttt sample.

%------------------------------ 
%%------------------------------ 

\section{New method to estimate $\rho$}
\label{sec:newRho}
In this section, we present the new method for $\rho$ estimation, which minimizes the dependence of $\rho$ on the presence of a hard-scatter event. The algorithm for this new method is based on the approach of the grid-median estimator, and it consists of the following steps:
\begin{enumerate}
\item Estimate the \yphi positions of clusters containing candidates for particles from hard-scatter events, \textit{signal seeds}. This step depends on the experimental environment, and there are three possibilities how signal seeds can be obtained:
\begin{enumerate}
\item Using all particles from an event-wide \pileup correction of the event. Typically, one can use the ICS or the CS method with Soft Killer. These combinations were previously studied e.g. in \cite{Berta:2019hnj,ATLAS:2017pfq}. The particles from \pileup-subtracted event are clustered into \akt jets obtained with jet radius $0.3-0.5$. A cut on minimum transverse momentum ($\pt$) or transverse momentum over the active jet area ($\pt/A$) is imposed to remove jets originating from \pileup due to imperfect \pileup correction. 
  % where \pt is the jet transverse momentum, and $A$ is the active jet area.
\item Using tracking information. A tracking detector can identify the origin of charged particles (hard-scatter event vs \pileup event) \cite{Sirunyan:2017ulk,Bertolini:2014bba}. The neutral particles can be pileup-corrected, and then the resulting neutral particles can be merged with charged hard-scatter particles to identify the signal seeds similarly as in case (a) (an example of using tracking information and correcting neutral particles is discussed in \secref{performance}). 
\item Signal seeds can also be given in the input manually as four-momenta. These can be e.g. four-momenta of jets previously identified by an external algorithm.
\end{enumerate}
\item Break the event into a grid of $N$ cells in the \yphi plane. Each cell has the same size~$a$.
\item Compute the momentum and area of each cell after excluding particles and areas close to signal seeds. The closeness is defined via a free parameter \Rexclude:
\begin{enumerate}
\item Transverse momentum of cell $i$, $\pti$, is defined as the scalar sum of $\pt$ of particles in the cell $i$.
Each particle $j$ has to satisfy the condition $\Delta R(\mathrm{particle}~j, \mathrm{seed}~k) > \Rexclude$ for any seed $k$, meaning that particles close to a seed are excluded from the sum.
\item Area of cell $i$, $A_i$, is defined as the geometric area in the \yphi plane for which each point fulfills again the condition $\Delta R(\mathrm{point}, \mathrm{seed}~k) > \Rexclude$ for any seed $k$. In practice, this geometric area can be estimated using ghost particles distributed with high density in the \yphi plane. 
\end{enumerate}
\item The estimate of $\rho$ is calculated as follows:
\begin{enumerate}
\item The densities of \pileup in cells, $\pti/A_i$, are ordered from the smallest to the largest, and a weight $w_i=A_i/\sum_i A_i$ is assigned to each density. The ordered densities and weights are labeled from 1 to $N$ as $\rho_1$, $\rho_2$,\ldots $\rho_N$, and $w_1$, $w_2$,\ldots $w_N$, respectively.
\item A function $f(w)$ is defined in range $w\in[0,1]$ as follows:

\begin{eqnarray}
\nonumber
f(w)=\rho_1~~~~~~& \mathrm{if} &~~~~  w\in[0,w_1) \\
\nonumber
f(w)=\rho_2~~~~~~& \mathrm{if} &~~~~  w\in[w_1,w_1+w_2) \\
f(w)=\rho_3~~~~~~& \mathrm{if} &~~~~ w\in[w_1+w_2,w_1+w_2+w_3) \\
\nonumber
& \ldots & \\
\nonumber
f(w)=\rho_N~~~~~~& \mathrm{if}& ~~~~  w\in[w_1+\ldots+w_{N-1},1]
\end{eqnarray}

\item The estimated $\rho$ is
\begin{equation}
\rhoest = \int\limits_{C-W/2}^{C+W/2} f(w) dw / W
\end{equation}
where $C$ and $W$ are free parameters controlling the center and width of the region used to compute $\rho$ from the grid cells, respectively.
\end{enumerate}
\end{enumerate}

\begin{figure}[h!]
  \centering
    \includegraphics[width=0.78\textwidth]{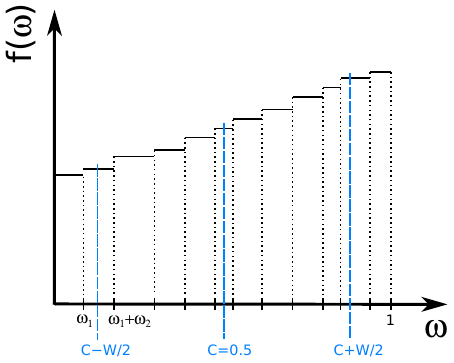}
    \caption{Cartoon illustrating the algorithm for calculating $\rhoest$.}
    \label{fig:drawing}
\end{figure}

A cartoon illustrating the algorithm for calculating $\rhoest$ is provided in \figref{drawing}.
The construction of function $f(w)$ is a generalization of the median calculation used in the grid-median method. The weights $w_i$ ensure that the contributions from the individual cells are proportional to the area of each cell. With $W<1$, outlying values of $\rho$ can be excluded. With $C<0.5$, the excluded region contains more cells with large $\rho$ and fewer cells with small $\rho$. The new method has two limits which return previously used methods for $\rho$ determination: 
\begin{itemize}
  \item Grid-median limit - When one sets $C=0.5$ and $W\rightarrow0$, the new method leads to a similar approach as in the grid-median method, just instead of standard median, a weighted median is used with weights $w_i$. When in addition no signal seeds are assigned, the new method is identical to the grid-median method.
  \item Average limit - When one sets $C=0.5$ and $W\rightarrow1$, then the estimated $\rho$ corresponds to the average. Effectively, it is the scalar sum of all non-excluded particles \pt divided by the non-excluded area. The effect of using a grid is entirely removed in this limit.
\end{itemize}

Besides setting parameter $C$ to be a fixed number, 
the possibility of having $C$ floating may also bring an advantage. In particular, the asymmetric configurations with $C<0.5$ floating depending on the total \pt of estimated hard-scatter particles may improve the performance of the $\rho$ determination, especially in the presence of a large number of jets outgoing the hard-scatter process. 
Therefore, following dependence of the $C$ parameter was implemented, 
\begin{equation}
C = 0.5-\min\left(\Smin,K\cdot H_\mathrm{T}/\sqrt{s}\right),
\end{equation}
where $H_\mathrm{T}$ is the scalar \pt sum of all the estimated hard-scatter particles, and $\sqrt{s}$ is the center-of-mass energy of the hadron collisions. The free parameter $K$ regulates how much the center $C$ depends on the hard-scatter activity. 
The free parameter \Smin represents the maximal allowed deviation of $C$ from 0.5. It is required to satisfy condition $0\leq\Smin\leq0.5-W/2$, where lower and upper bounds lead to $C=0.5$ and $C=W/2$, respectively.

The same rapidity-rescaling approach as for the grid-median method described by \eref{rapidity_rescaling} can be used in the new method. The rapidity rescaling is not discussed in the above description of the new method for clarity, but it is implemented in the software implementation of \mname~in \texttt{FastJet Contrib}.

\section{Performance}
\label{sec:performance}

\subsection{Performance of $\rho$ determination}
\label{sec:rho}

To evaluate the performance of the new method for $\rho$ estimation, we tested various configurations. We present two representative configurations that give significantly better performance than the grid-median method: 
\begin{itemize}
 \item Large grid size with fixed $C$. The parameters used are $a=1.2$, $C=0.5$, and $W=0.2$.
 \item Small grid size with floating $C$. The used parameters are: $a=0.55$, floating $C$ with $K=0.1$ and $\Smin=0.1$, and $W=0.8$.
\end{itemize}
For both configurations, the signal seeds are found using the approach of step 1 (a) in \secref{newRho}, i.e. the tracking information is not used. The event is first corrected with ICS method with two iterations (maximal distance parameters are 0.2 and 0.35) and using the grid-median method to estimate $\rho$ (cell size of 0.55). This pileup corrected event is clustered with the \akt jet clustering algorithm with the distance parameter of 0.3. The signal seeds are then defined as the clustered jets fulfilling the condition $\pt/A>(p_0+p_1 \cdot\sqrt{\rho/\gev})$ 
where $p_0,p_1$ are free parameters, \pt is the jet transverse momentum, $A$ is the active jet area, and $\rho$ is the estimated pileup \pt density in the event using the grid-median approach at rapidity $y=0$. This $\rho$-dependent threshold is used as a compromise between minimizing the number of pileup jets to be defined as signal seeds and maximizing the number of signal seeds originating from the hard-scatter event. In MC studies using previously defined samples, we found optimal values of free parameters to be $p_0=5\gev$ and $p_1=8\gev$. Further, for both configurations, the exclusion radius $\Rexclude$ is set to 0.4, which is larger than the distance parameter of the \akt algorithm to ensure that all hard-scatter particles are excluded.

\begin{figure}[!h]
  \centering
  \begin{subfigure}{0.48\textwidth}
    \includegraphics[width=\textwidth]{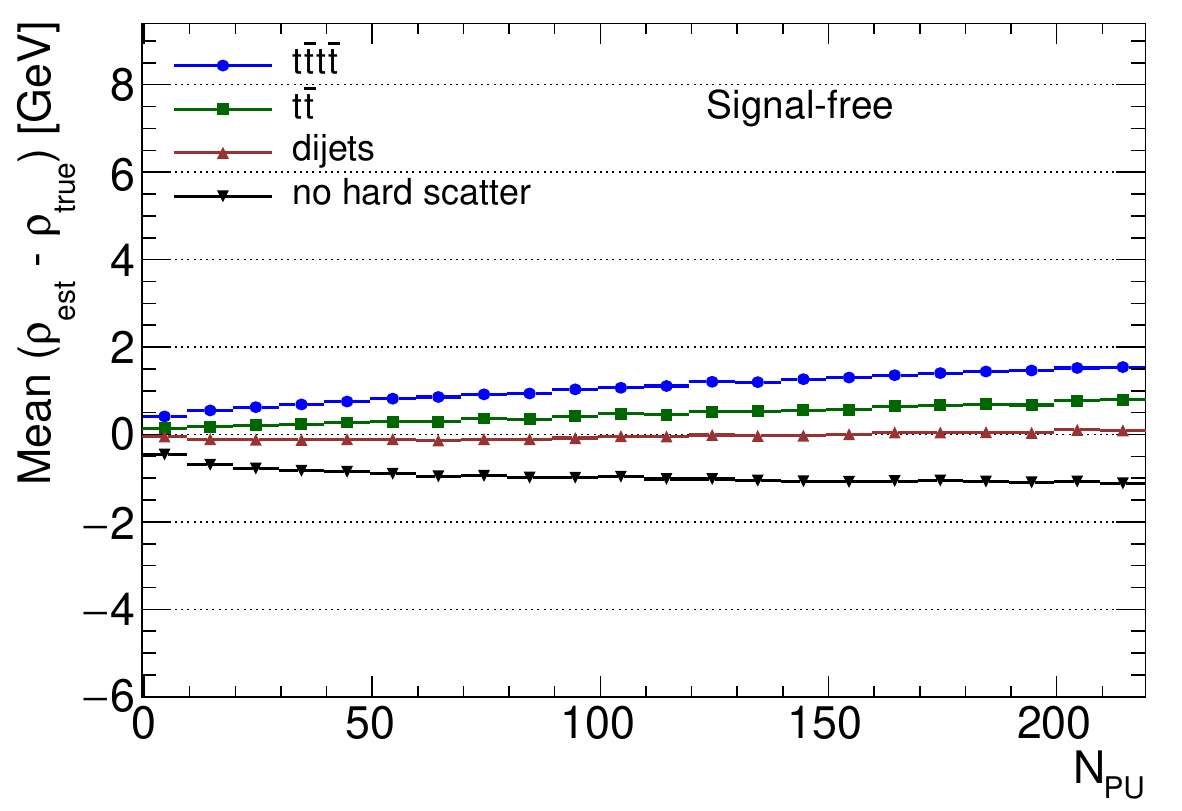}
    \caption{$a$=1.2, $W=0.2$, $C=0.5$}
    \label{fig:new_W0_mean}
  \end{subfigure}
  \begin{subfigure}{0.48\textwidth}
    \includegraphics[width=\textwidth]{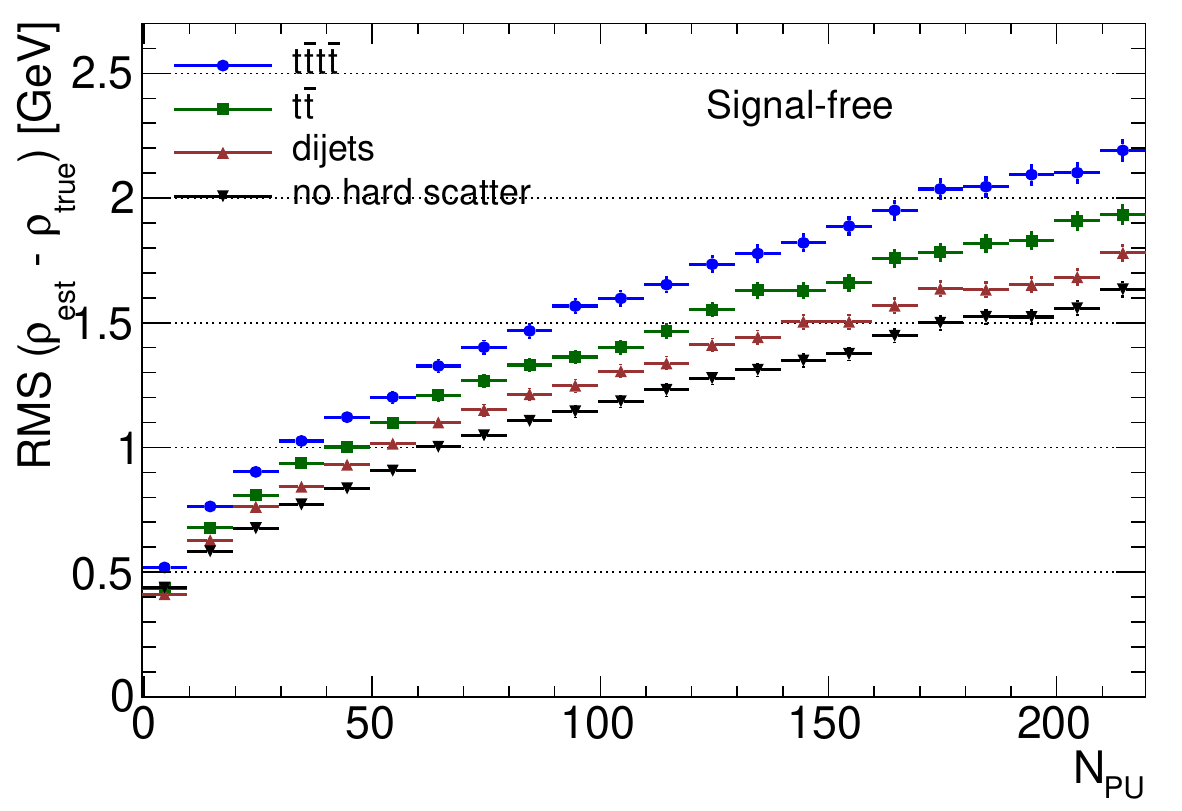}
    \caption{$a$=1.2, $W=0.2$, $C=0.5$}
    \label{fig:new_W0_rms}
  \end{subfigure}
  \begin{subfigure}{0.48\textwidth}
    \includegraphics[width=\textwidth]{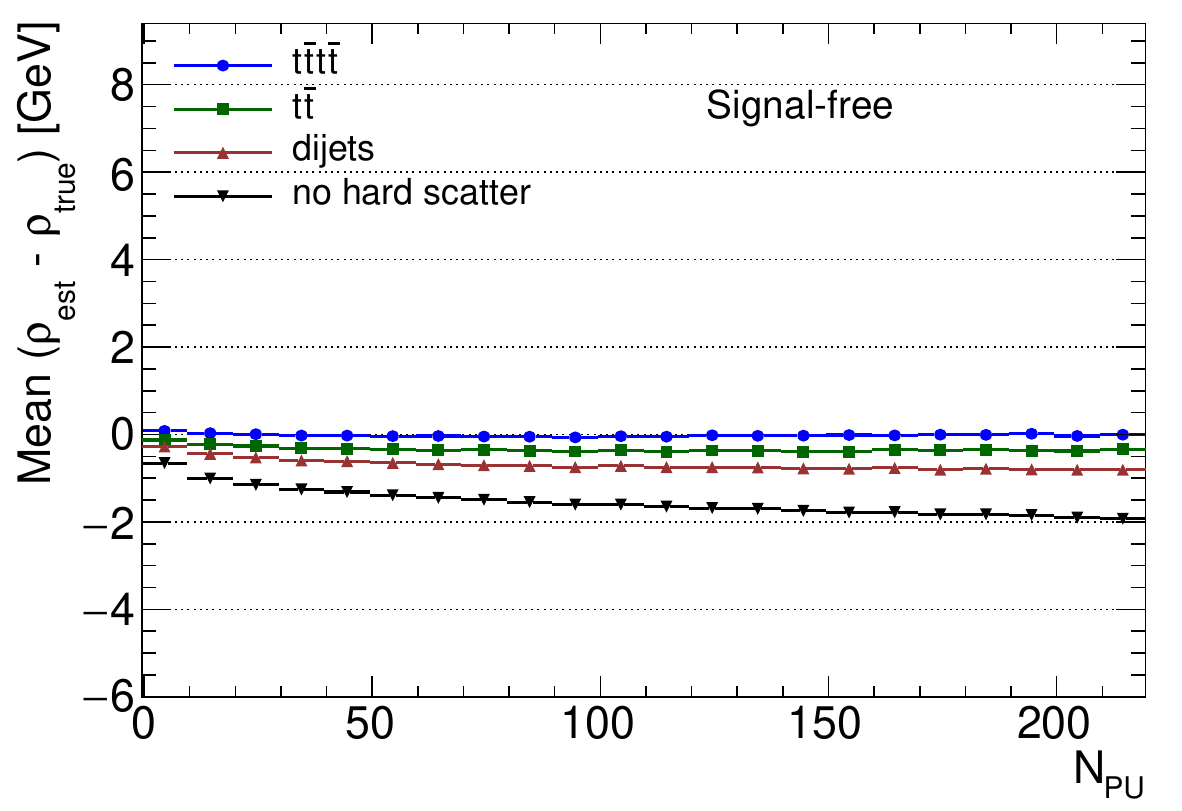}
    \caption{$a$=0.55, $W=0.8$, floating $C$}
    \label{fig:new_W0.8_mean}
  \end{subfigure}
  \begin{subfigure}{0.48\textwidth}
    \includegraphics[width=\textwidth]{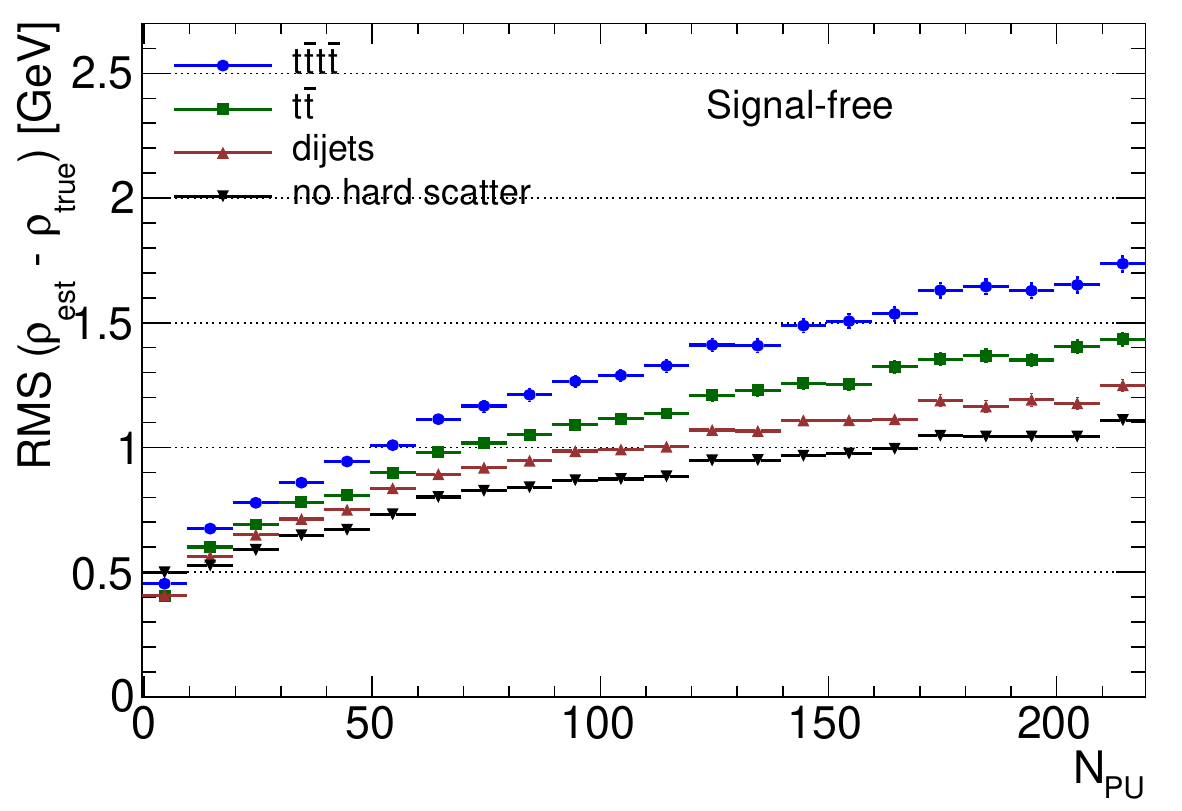}
    \caption{$a$=0.55, $W=0.8$, floating $C$}
    \label{fig:new_W0.8_rms}
  \end{subfigure}
  \caption{Dependence of $\rho$ bias (left) and resolution (right) on \npu for $\rho$ obtained using the new method (here labeled "Signal-free", for details see the text) without access to the information about the charged particles. The dependence is shown for four samples: no hard-scatter process, dijets, \ttbar production, and four top quark production.}  
  \label{fig:newMethod}
\end{figure}

\Figref{newMethod} quantifies the performance in terms of the $\rho$ bias and resolution evaluated as a function of \npu for four samples (no hard-scatter process, dijets, \ttbar production, and \tttt production). Samples are identical to those used in \secref{stateOfTheArt} and $\rho$ is again evaluated at rapidity $y = 0$.  One can see that the spread in the $\rho$ bias decreased significantly for both configurations. When the grid-median method with recommended settings was used, the spread in $\rho$ bias at $\npu = 60$ was 4 GeV. With the new method, the spread is ${\sim}1.5\gev$ at $\npu = 60$ and stays below 2 GeV even at the maximum \npu of 220. The remaining \npu dependence of $\rho$ bias is milder for all samples and is approximately linear. While the performance in terms of bias is similar for both configurations, the resolution of $\rho$ is significantly improved for all the samples in the case of $a=0.55$, $W=0.8$ and floating $C$. The improvement is the most pronounced in the dijet sample and in the sample with no hard scattering when the resolution improves by about 50\%.

The $\rho$ estimate may be significantly further improved when using the information from tracking (step 1 (b) in \secref{newRho}). First, we can assume that the experiment can reconstruct jets from tracks connected with the hard scatter using vertexing. The charged hard-scatter particles are clustered using the \akt jet clustering algorithm with the distance parameter of 0.3. The signal seeds are then defined as the clustered jets fulfilling the condition $\pt/A>20\gev$ (no $\rho$-dependent threshold is needed here since these jets are not affected by the pileup). 
To include also the neutral component a second set of seeds is defined. These seeds are obtained by clustering the pileup-corrected event and applying the $\rho$-dependent $\pt/A$ cut defined earlier. In this case, the pileup-corrected event is obtained by applying the charged hadron subtraction method \cite{CMS-PAS-JME-14-001} followed by ICS. The other parameters of the new method are set to be: $a=1.2$, $\Rexclude=0.4$, floating $C$ with $K=0.1$, and $\Smin=0.1$, $W=0.8$.

\begin{figure}[!h]
  \centering
  \begin{subfigure}{0.48\textwidth}
    \includegraphics[width=\textwidth]{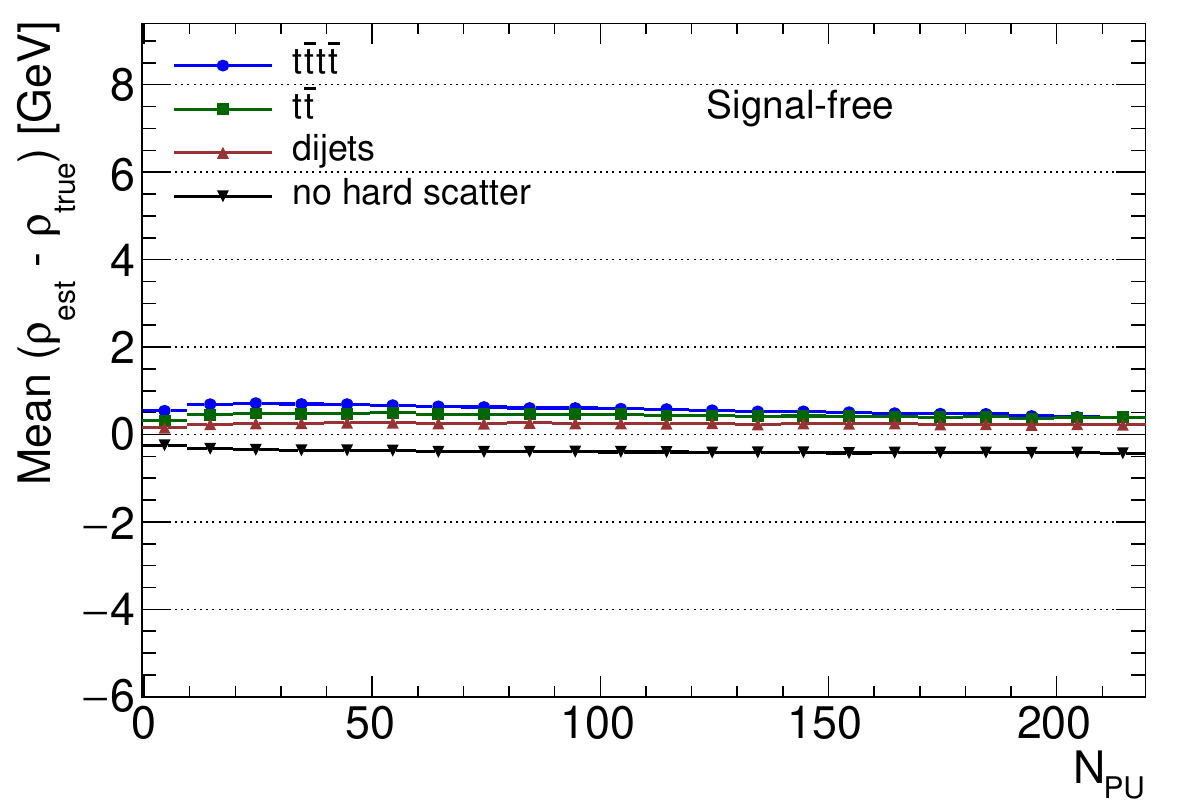}
    \label{fig:new_charged_W0.8_mean}
  \end{subfigure}
  \begin{subfigure}{0.48\textwidth}
    \includegraphics[width=\textwidth]{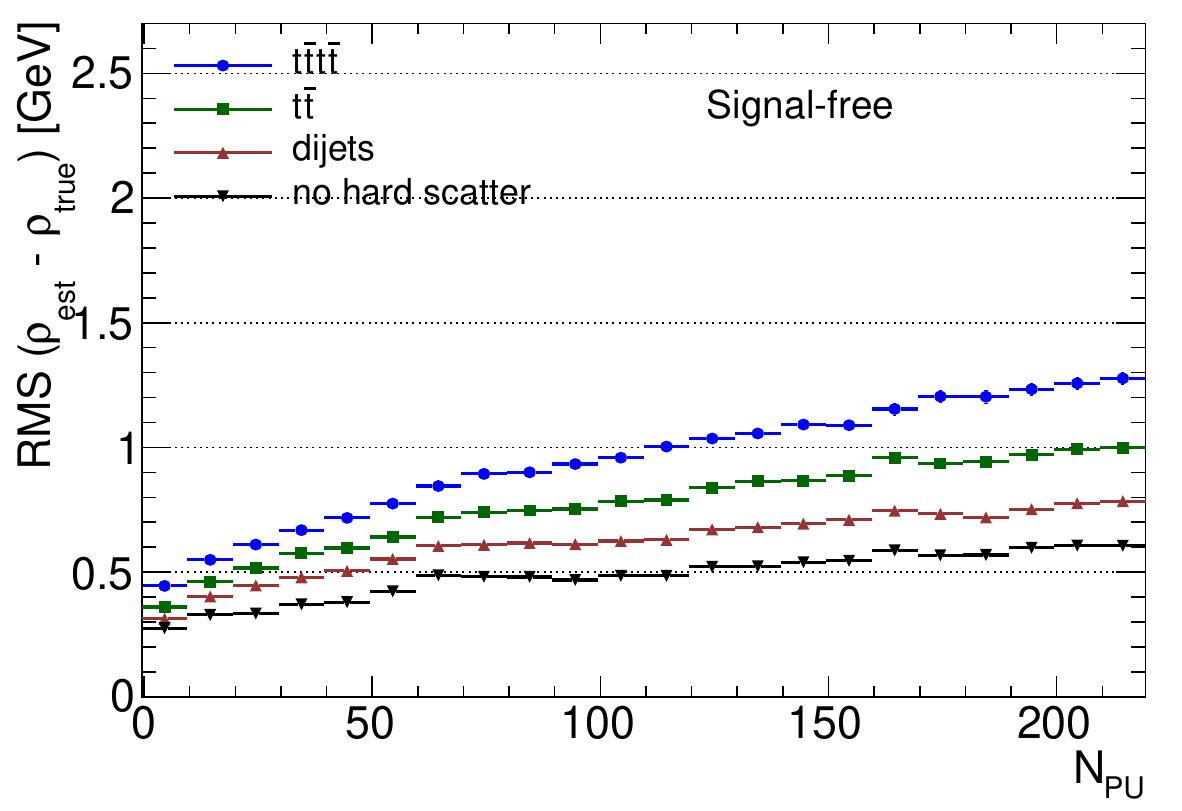}
    \label{fig:new_charged_W0.8_rms}
  \end{subfigure}
  \caption{Dependence of $\rho$ bias (left) and resolution (right) on \npu for $\rho$ obtained using the new method with the access to the information about the charged particles (here labeled "Signal-free", for details see the text). The dependence is shown for four samples: no hard-scatter process, dijets, \ttbar production, and 4 top quark production.}  
  \label{fig:newMethod_charged}
\end{figure}

The $\rho$ bias and resolution after using the information from tracking are shown in \figref{newMethod_charged}. One can see that when using the tracking information, the \npu dependence was practically removed. The spread in $\rho$ bias is smaller than 1 GeV for all the \npu values, and the resolution improved significantly. 
This result implies that if this performance can be maintained in the realistic detector, no additional pileup-dependent calibrations of the jet energy scale are needed. Moreover, reducing the spread in $\rho$ bias and improving the $\rho$ resolution should reduce systematic uncertainties in measurements of observables that are not fully unfolded for the impact of resolutions induced by the pileup. This is important for current as well as future precision measurements.

An improvement of $\rho$ estimation can be achieved not only by using the tracking information but also by repeating the new method iteratively twice. In that case, the pileup subtraction used to define the signal seeds in the second iteration uses the estimated $\rho$ from the first iteration.

Several parameter choices were made in this section which were determined to deliver the best performance for a given setup. The choice of optimal parameters is expected to depend highly on the experimental setup and it can therefore be different from choices made here. A discussion of the impact of the choice of parameters on the $\rho$ estimation as well as the impact of individual steps in the new algorithm is provided in \appref{parameters}.

%%------------------------------ 

\subsection{Performance of jet reconstruction}
\label{sec:jet}
To quantify the performance of the new method, we evaluate the subtracted jet spectra and compare them with the jet spectra of events with the same hard process but with no pileup added. The background subtraction is done using jet-by-jet CS. In general, the performance of any observable can be evaluated in terms of the \textit{bias} and \textit{resolution} of a given observable, $x$. These quantities can be defined as follows:

  \begin{eqnarray}
  \mathrm{bias} = \frac{\langle x^{\mathrm{rec}} - x^{\mathrm{true}}\rangle}{\langle x^{\mathrm{true}}\rangle}, ~~~
  \mathrm{resolution} = \frac{\mathrm{RMS}( x^{\mathrm{rec}} - x^{\mathrm{true}})}{\langle x^{\mathrm{true}}\rangle}.
  \end{eqnarray}

The difference between $x^{\mathrm{rec}}$ and $x^{\mathrm{true}}$ is used instead of their ratio to avoid excessive values of bias for small values of $x^{\mathrm{true}}$. The denominator in the bias and resolution allows for easier comparison among different quantities. In the case of the jet spectra, the bias and resolution are called jet energy scale and jet energy resolution, respectively. 

\begin{figure}[!h]
  \centering
  \begin{subfigure}{0.48\textwidth}
    \includegraphics[width=0.96\textwidth]{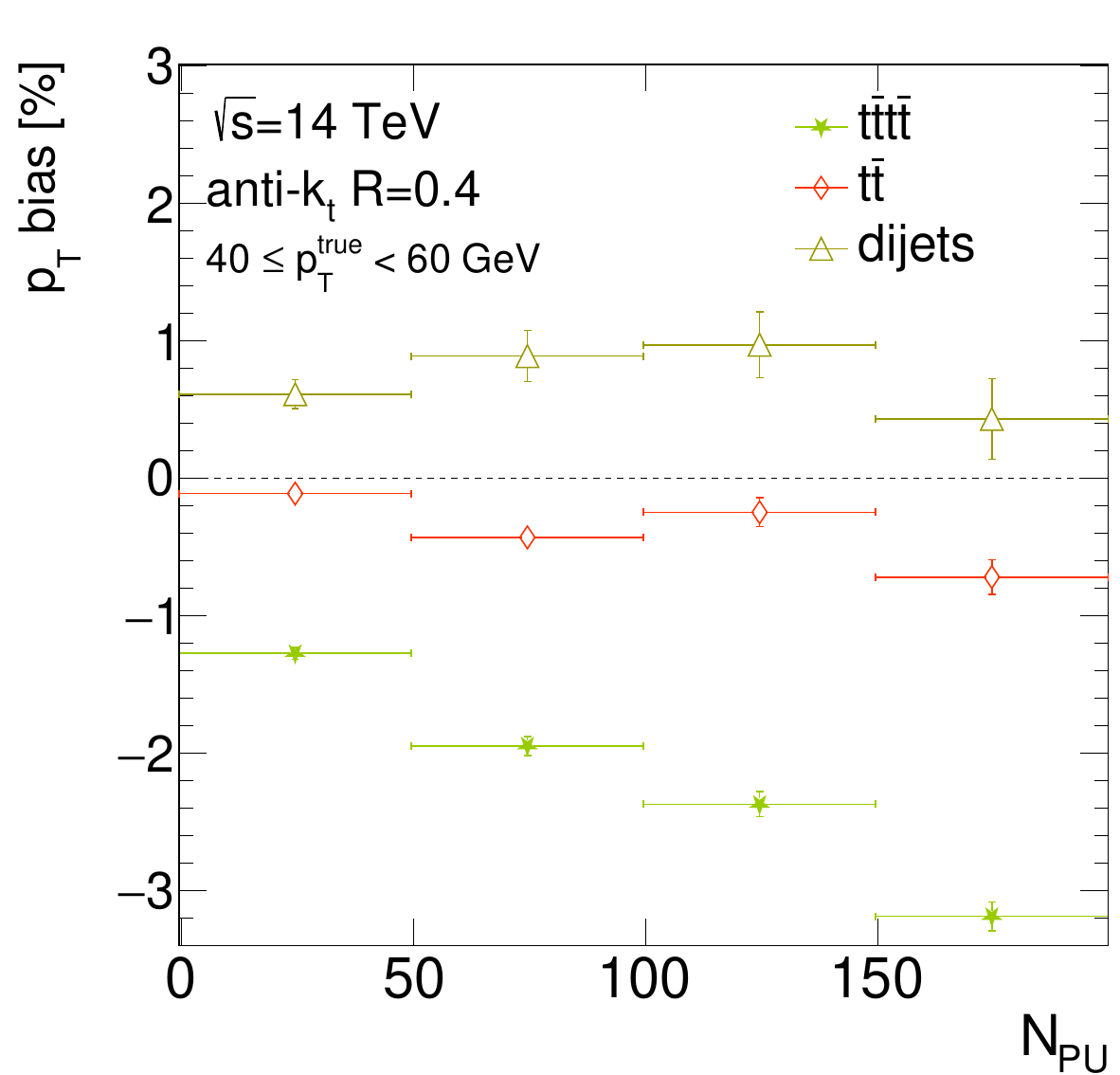}
    \caption{grid-median size 0.55}
    \label{fig:CS_grid}
  \end{subfigure}
  \begin{subfigure}{0.48\textwidth}
    \includegraphics[width=0.96\textwidth]{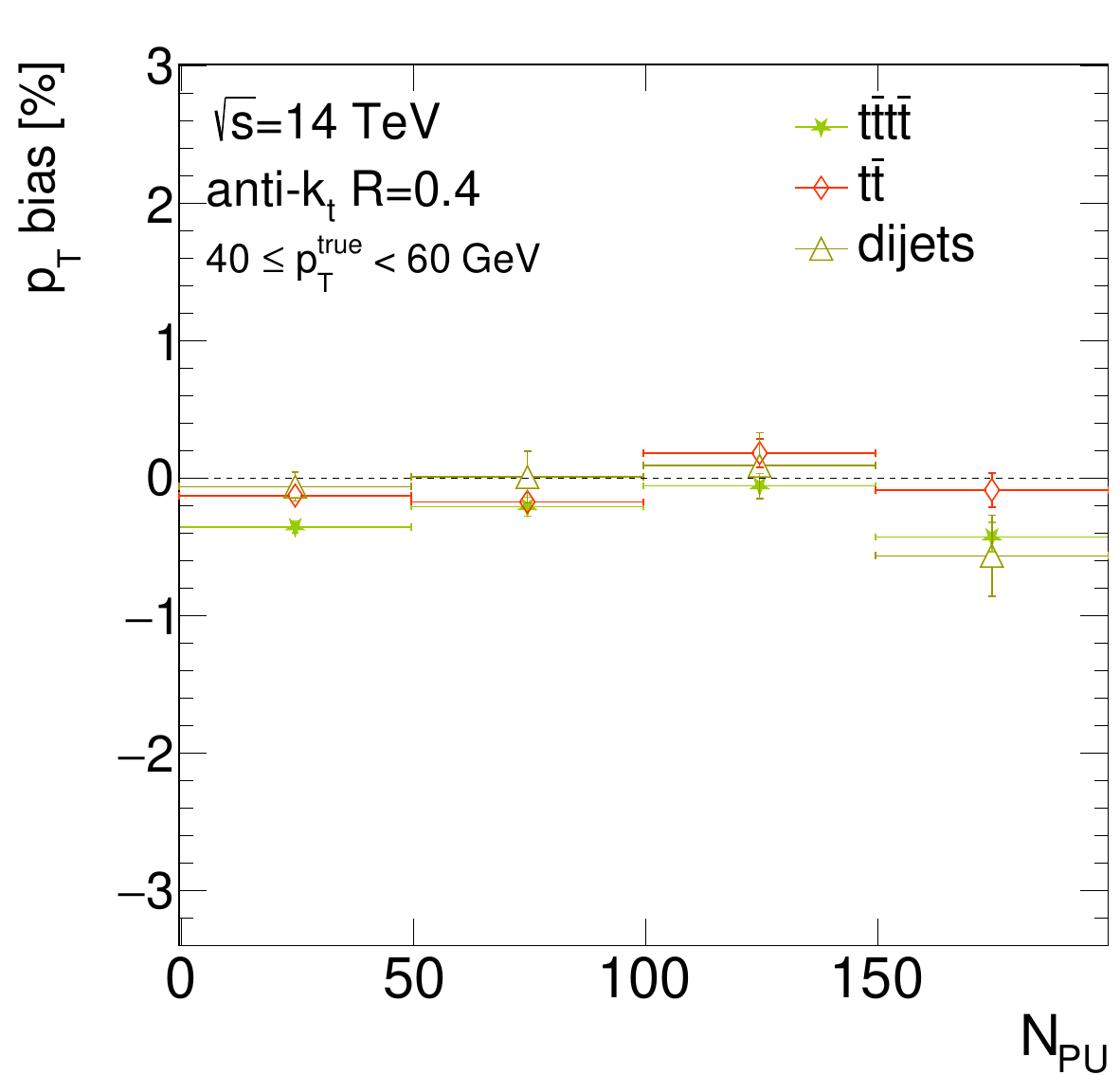}
    \caption{new method}
    \label{fig:CS_new}
  \end{subfigure}
  \caption{Performance of pileup correction for jets using the jet-by-jet Constituent Subtraction method with the grid-median method with recommended settings (left) and the new method using tracking information (right). The jet \pt bias is plotted as a function of \npu for \akt $R=0.4$ jets with true $\pt\in[40,60]\gev$. The dependence is shown for three signal samples: dijets (yellow), \ttbar production (red), and four top quark production (green).}  
  \label{fig:jet_pt}
\end{figure}

\Figref{jet_pt} shows the jet energy scale evaluated as a function of \npu for jets subtracted using CS with two different input $\rho$ values. In the first case, the $\rho$ is estimated using the standard grid-median approach with grid size $a=0.55$. In the second case, the $\rho$ is estimated using the new method using tracking information as described in \secref{rho}. One can see that in the case of using the standard method, the jet energy scale differs from zero by 1-3\%. This difference depends on the sample and it is the largest for the sample with four top quarks where the final state jet multiplicity is the largest. In the case of using the new method, the jet energy scale is smaller than 0.5\% for all samples and for all \npu values. This represents a significant improvement in the ability to reconstruct final state observables, which should clearly impact the precision of measurements involving jets reconstructed in the noisy environment of high pile-up events.

Comparison of the jet performance of the new method with the performance of a method independent on $\rho$ is provided in \appref{framework_pileup_workshop}.

\section{Conclusions}
\label{sec:conclusions}
We presented a new method to estimate the pileup density, $\rho$, which is an essential ingredient for multiple state-of-the-art pileup mitigation methods. It is based on excluding areas in the \yphi plane containing jets from the hard-scatter event and on a generalized approach to compute the median from grid cells. We showed that the new method minimizes the dependence on the number of hard-scatter jets in the event. 
The difference between the true $\rho$ estimated using purely the pileup event and the $\rho$ estimated in the presence of hard-scatter events using the new method is smaller than 1 GeV if the information about charged tracks is used. This difference is \npu independent, and it translates to a negligible impact of pileup on the jet energy scale.
The new method has therefore the potential to significantly improve the pileup mitigation in proton-proton collisions or the underlying event subtraction in heavy-ion collisions.

\appendix

\acknowledgments
This project has received funding from the PRIMUS research programme of Charles University (project No 21/SCI/017), from Grant Agency of the Czech Republic under grant 22-11846S, from project ERC-CZ LL2327, and from Charles University Research Centre programs UNCE/SCI/013 and UNCE/24/SCI/016.

\section{Choice of parameters for the new method}
\label{app:parameters}
In \secref{rho}, the performance of the new method for the background estimate was documented. A significant improvement over the methods with standard $\rho$ estimate was seen. It may also be useful to quantify the impact of parameter choice and individual steps in the new $\rho$ algorithm, as described in \secref{newRho}, on this performance. Such a quantification is provided here.

\begin{figure}[!h]
  \centering
    \begin{subfigure}{0.41\textwidth}
    \includegraphics[width=\textwidth]{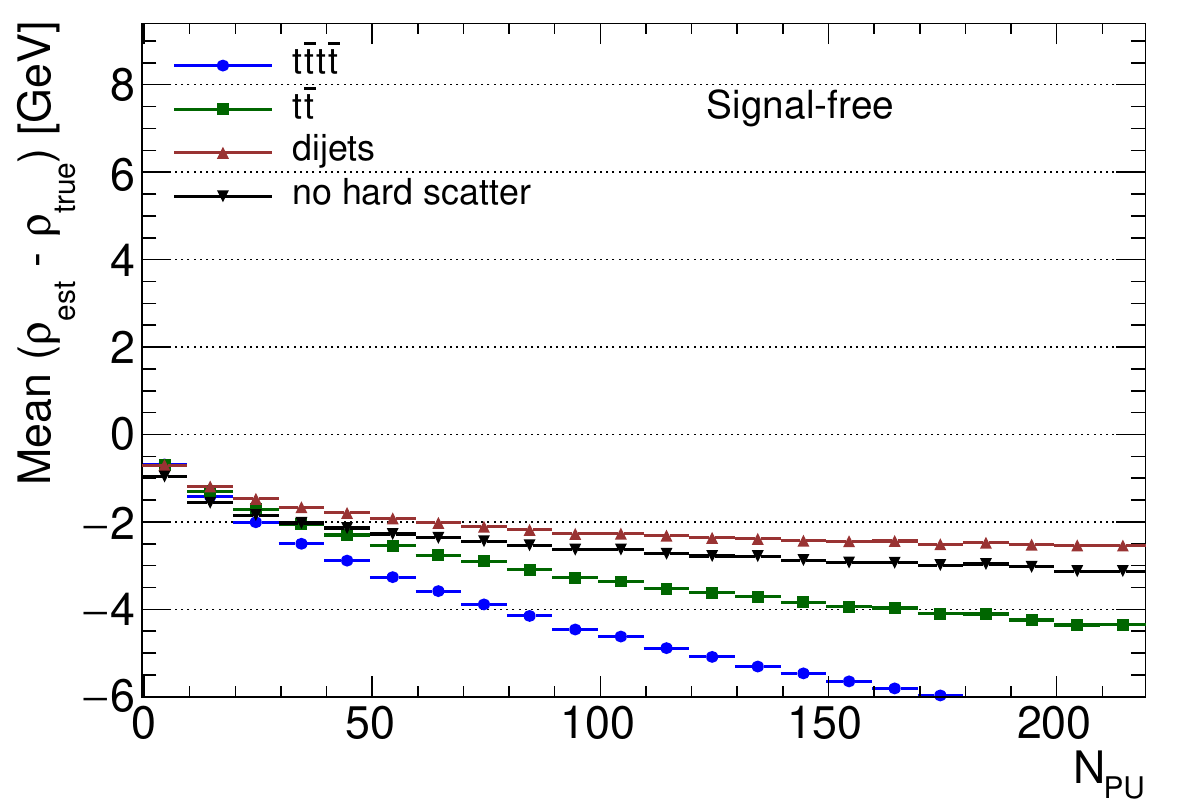}
    \caption{$a$=0.55, Steps 1-3}
    \label{fig:new_steps1-3_mean}
  \end{subfigure}
  \begin{subfigure}{0.41\textwidth}
    \includegraphics[width=\textwidth]{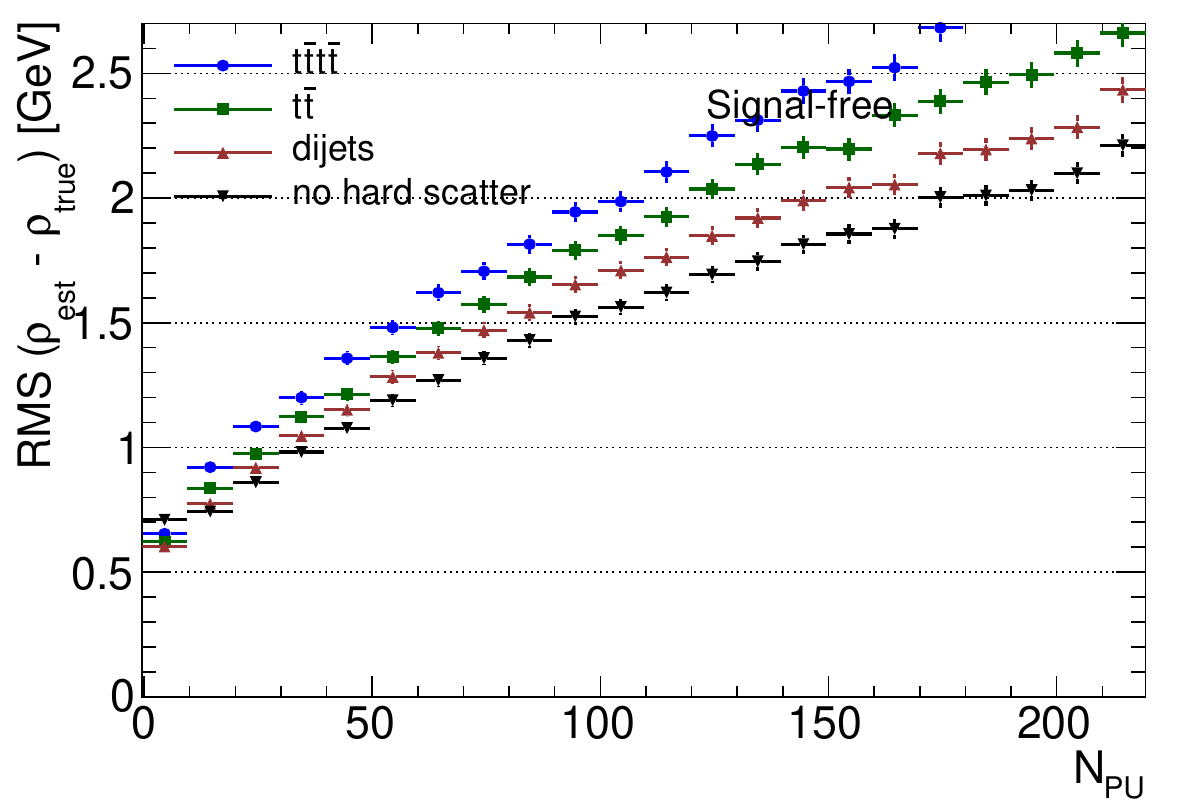}
    \caption{$a$=0.55, Steps 1-3}
    \label{fig:new_steps1-3_rms}
  \end{subfigure}
  \begin{subfigure}{0.41\textwidth}
    \includegraphics[width=\textwidth]{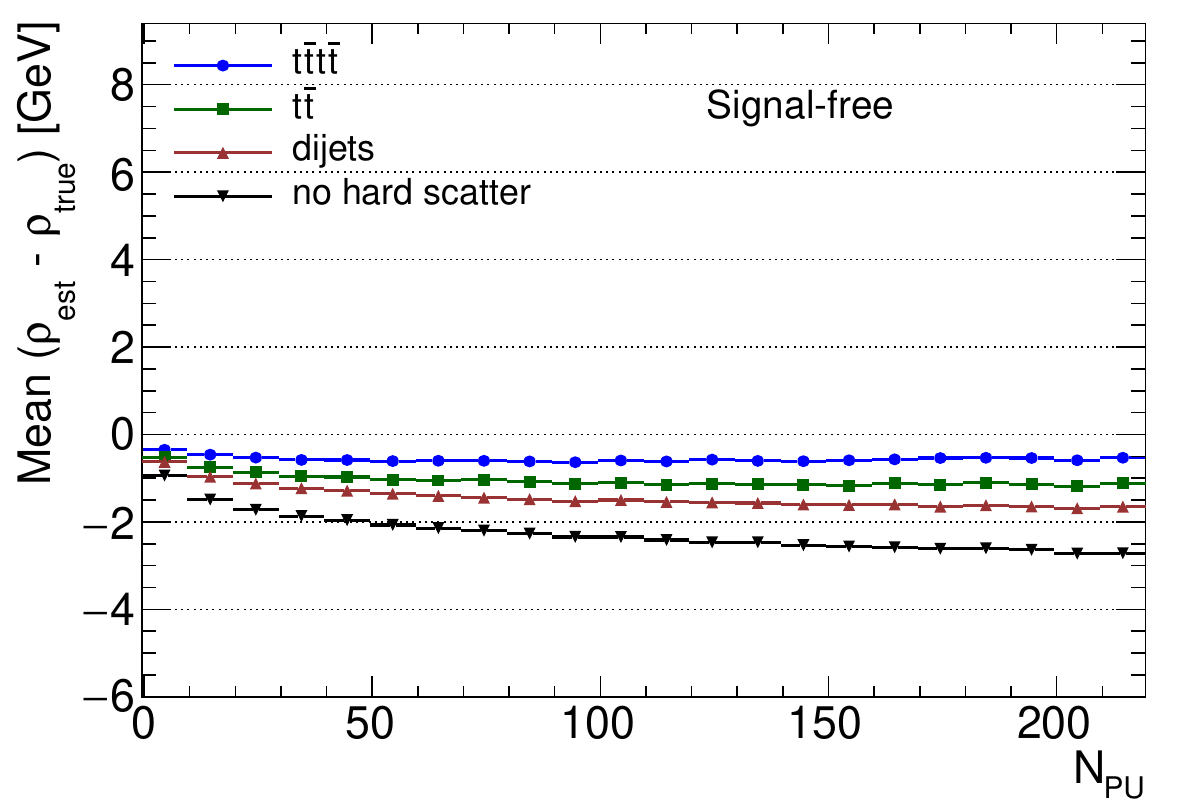}
    \caption{$a$=0.55, $W\rightarrow0$, fixed $C=0.5$}
    \label{fig:new_W0_fixedC_mean}
  \end{subfigure}
  \begin{subfigure}{0.41\textwidth}
    \includegraphics[width=\textwidth]{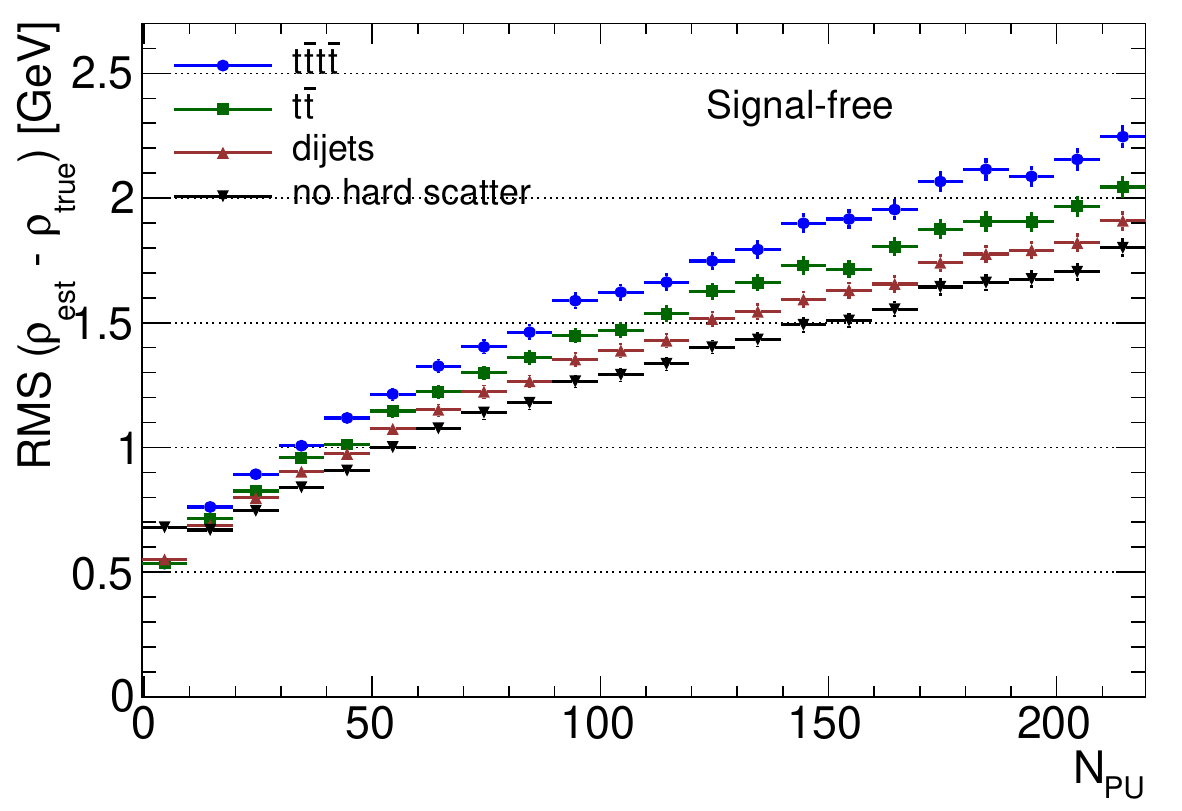}
    \caption{$a$=0.55, $W\rightarrow0$, fixed $C=0.5$}
    \label{fig:new_W0_fixedC_rms}
  \end{subfigure}
  \begin{subfigure}{0.41\textwidth}
    \includegraphics[width=\textwidth]{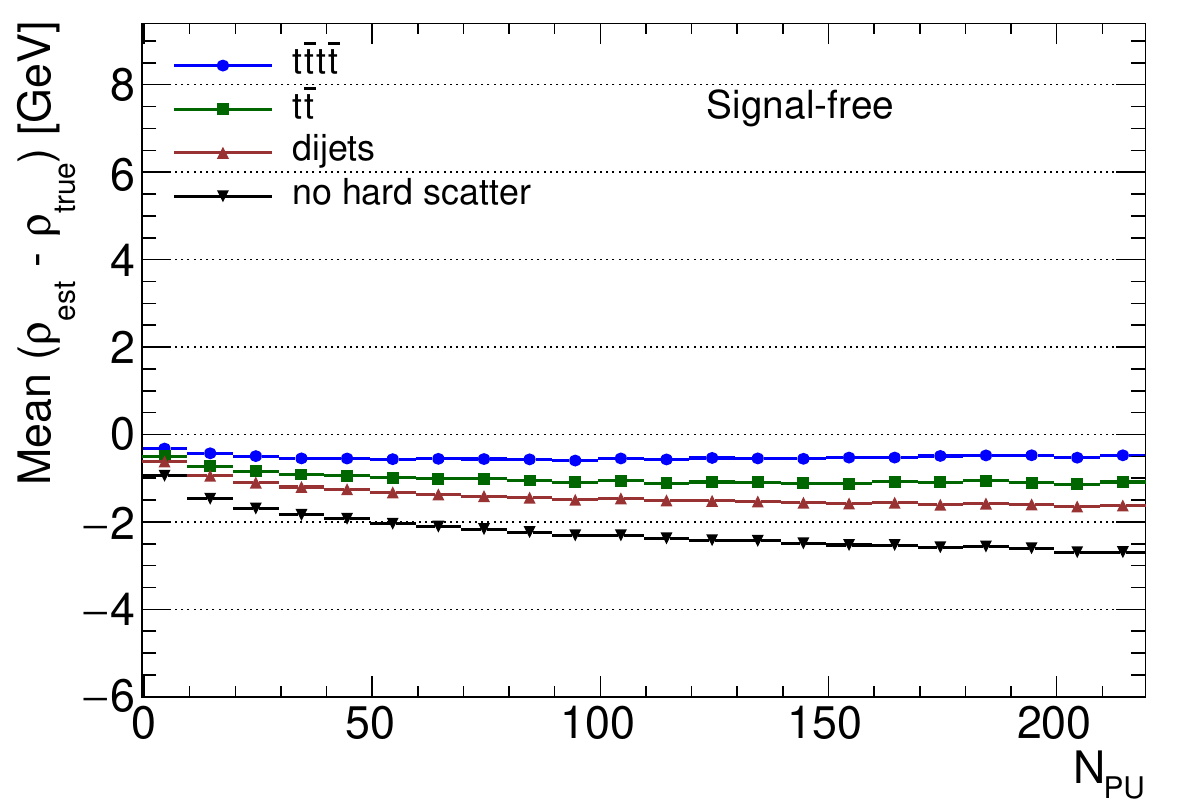}
    \caption{$a$=0.55, $W=0.2$, fixed $C=0.5$}
    \label{fig:new_W0.2_fixedC_mean}
  \end{subfigure}
  \begin{subfigure}{0.41\textwidth}
    \includegraphics[width=\textwidth]{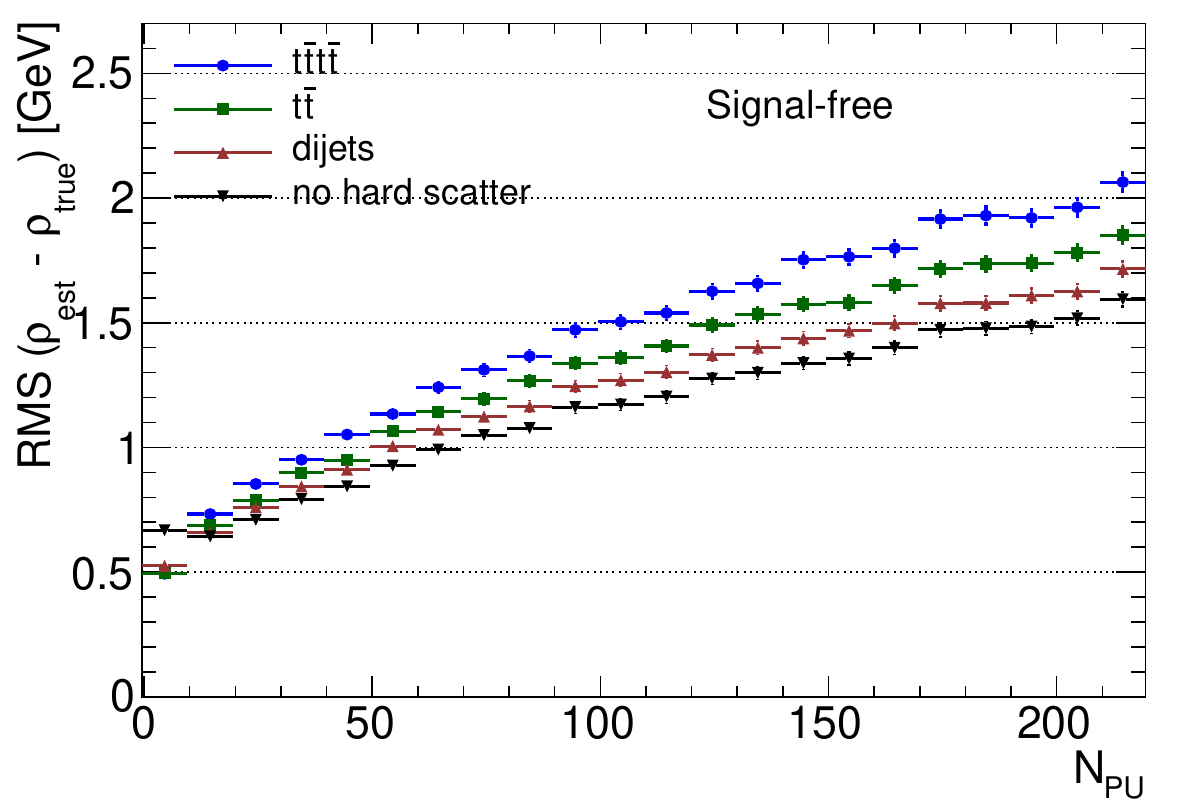}
    \caption{$a$=0.55, $W=0.2$, fixed $C=0.5$}
    \label{fig:new_W0.2_fixedC_rms}
  \end{subfigure}
    \begin{subfigure}{0.41\textwidth}
    \includegraphics[width=\textwidth]{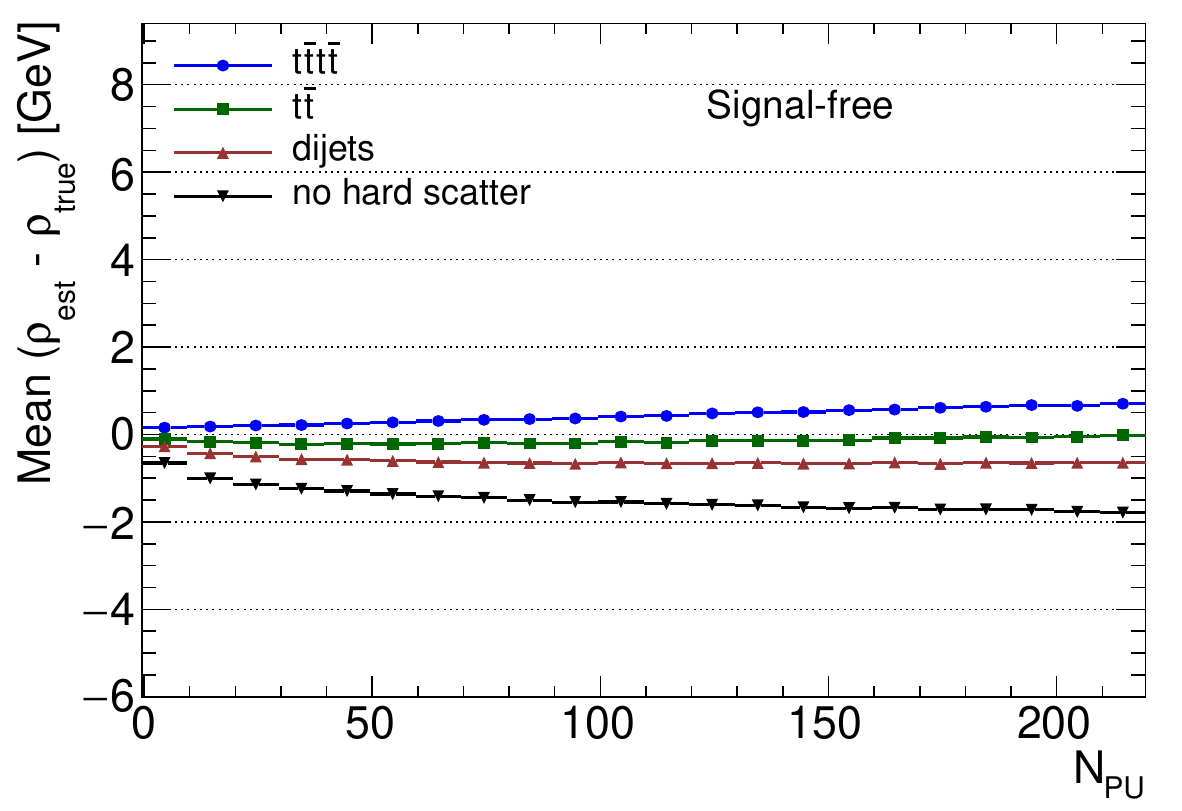}
    \caption{$a$=0.55, $W=0.8$, fixed $C=0.5$}
    \label{fig:new_W0.8_fixedC_mean}
  \end{subfigure}
  \begin{subfigure}{0.41\textwidth}
    \includegraphics[width=\textwidth]{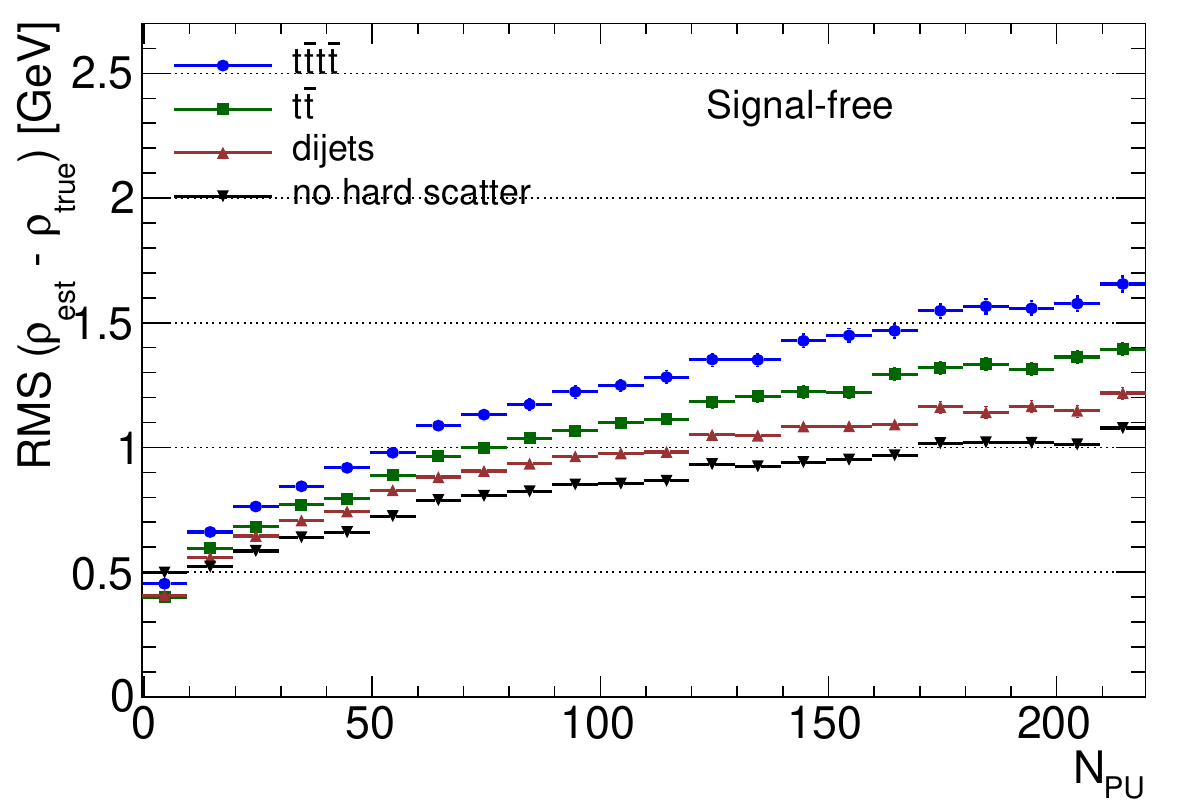}
    \caption{$a$=0.55, $W=0.8$, fixed $C=0.5$}
    \label{fig:new_W0.8_fixedC_rms}
  \end{subfigure}
  \caption{Dependence of $\rho$ bias (left) and resolution (right) on \npu for $\rho$ obtained using the new method for various parameters of the new method. The dependence is shown for four samples: no hard-scatter process, dijets, \ttbar production, and 4 top quark production.}  
  \label{fig:newMethod_parameters}
\end{figure}

\Figref{newMethod_parameters} shows bias and resolution of $\rho$ (defined in \secref{current_performance}) when performing only certain steps of the new algorithm and when using specific configuration to determine $\rho$. \Figsref{new_steps1-3_mean}{new_steps1-3_rms} shows the performance when areas around signal seeds are covered and subsequently running the standard grid-median approach to calculate $\rho$ (i.e. only Steps $1-3$) with grid size $a=0.55$. The calculation of $\rho$ does not use the areas of cells, $A_i$, as weights in this case. The signal seeds are obtained from the pileup-corrected events without using tracking information in the same way as in \secref{rho}, i.e., using parameters $p_0=5\gev$ and $p_1=8\gev$. The $\rho$ is highly underestimated for hard-scatter events with many jets in this configuration, and hence the performance is not better than for the grid-median method. An improved performance can be achieved by increasing the threshold for signal seeds via increasing parameters $p_0$ and $p_1$. In the limit $p_0\rightarrow\infty$ or $p_1\rightarrow\infty$, no signal seeds are defined meaning the new method is equivalent to the grid-median method.

The above configuration can be modified to use the areas of cells as weights when computing the median (this is equivalent to the full signal-free method using fixed $C=0.5$ and $W\rightarrow0$). The performance of this configuration is shown in \figsref{new_W0_fixedC_mean}{new_W0_fixedC_rms}. The bias and resolution are much better in this configuration, which clearly demonstrates the need to use the cell areas, $A_i$, as weights in the computation of the median. The results in \figsref{new_W0_fixedC_mean}{new_W0_fixedC_rms} can also be compared with results shown in \secref{current_performance}, quantifying the performance of state-of-the-art methods. One can see a clear improvement in the bias of $\rho$ even after performing only the first three steps of the new algorithm and using a weighted median. 

\Figrange{new_W0.2_fixedC_mean}{new_W0.8_fixedC_rms} then show the performance of the new method using the same configuration as above, but with two higher sizes of the window: $W=0.2$ and $W=0.8$. The center $C$ is fixed at $C=0.5$. While the bias is slightly improved with increasing $W$, a significant improvement is achieved in the resolution for all the used samples. The presented configuration with $W=0.8$ is almost identical to the one called "Small grid size with floating $C$" in \secref{rho} with one difference: In the latter configuration, a floating center is also used instead of a fixed center at 0.5. Therefore, by comparing \figsref{new_W0.8_fixedC_mean}{new_W0.8_fixedC_rms} with \figsref{new_W0.8_mean}{new_W0.8_rms}, one can see the impact of using the floating center $C$. Using floating $C$, reduces the differences in bias among the individual hard-scatter samples.

In general, these results show that the choice of the window in Step 4 significantly impacts performance. The comparisons presented here should help the reader become sensitive to the parameter settings, which can then be optimized in a given experimental environment to obtain the best possible performance.

\begin{figure}[!h]
  \centering
  \begin{subfigure}{0.48\textwidth}
    \includegraphics[width=\textwidth]{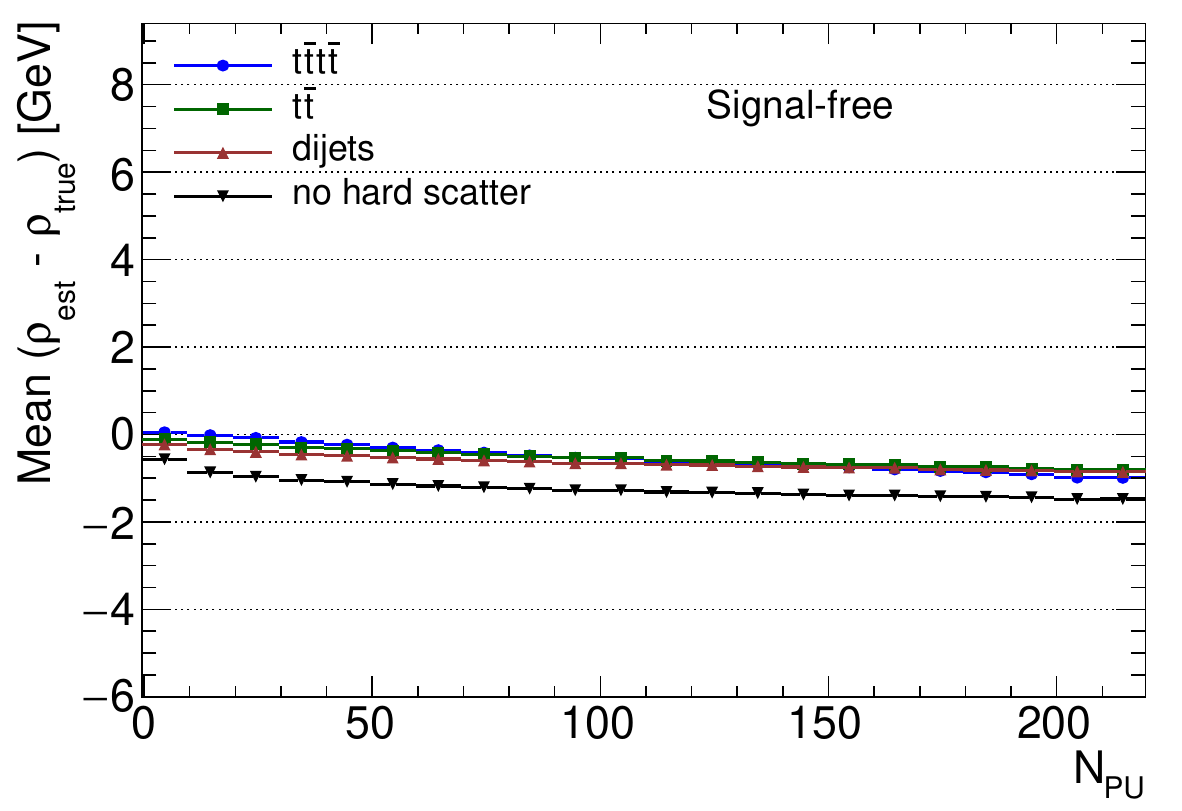}
    \label{fig:new_charged_W0.4_a0.55_mean}
  \end{subfigure}
  \begin{subfigure}{0.48\textwidth}
    \includegraphics[width=\textwidth]{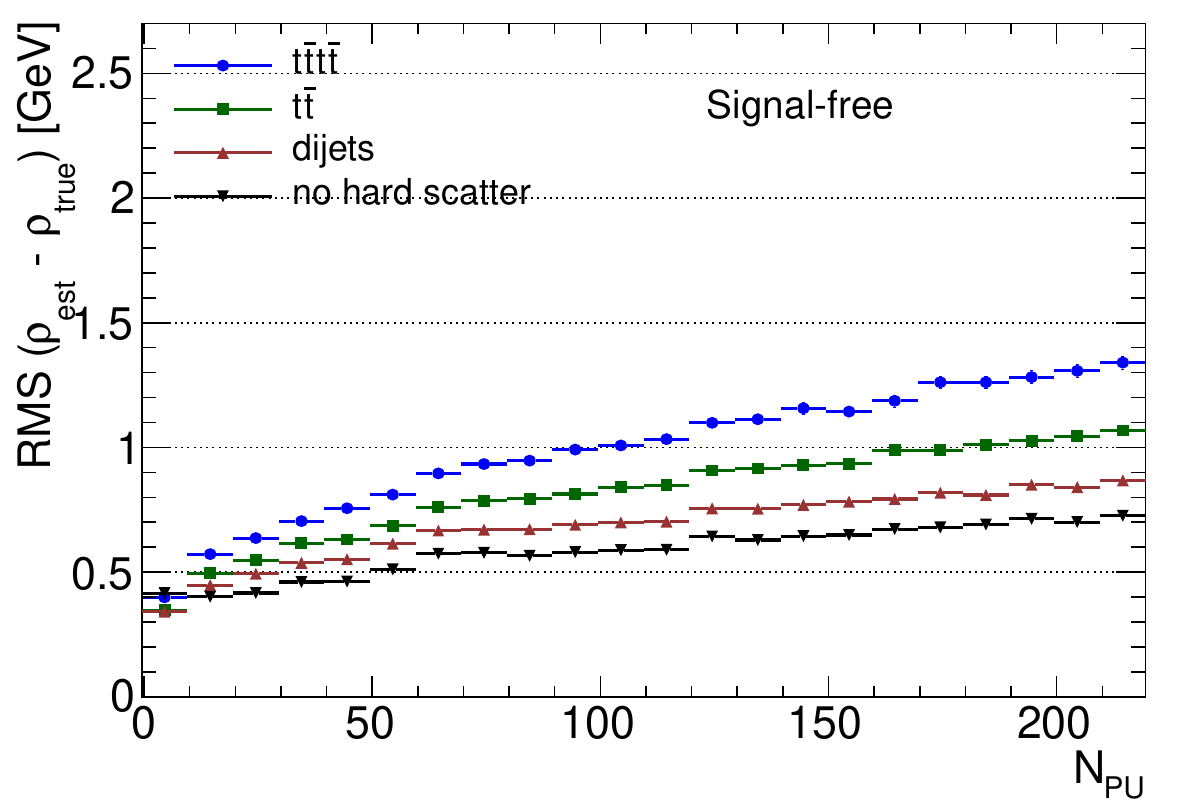}
    \label{fig:new_charged_W0.4_a0.55_rms}
  \end{subfigure}
  \caption{Dependence of $\rho$ bias (left) and resolution (right) on \npu for $\rho$ obtained using the new method with the access to the information about the charged particles. The dependence is shown for four samples: no hard-scatter process, dijets, \ttbar production, and 4 top quark production. Grid size parameter $a=0.55$.}  
  \label{fig:newMethod_charged_a0.55}
\end{figure}

To document how the improvement in the $\rho$ estimation achieved when using the tracking information (\figref{newMethod_charged}) is impacted by a choice of $a$ parameter, we show in \figref{newMethod_charged_a0.55} the performance for $a=0.55$ instead of $a=1.2$. The differences in the bias among samples are comparably small as in the case of using $a=1.2$, but the bias is, in general, more shifted from zero, and the resolution is worse by about 10\%. The slightly better performance of $a=1.2$ case might be related to the fact that increasing $a$ implies that $\rho$ is closer to the average \pt value per unit area than to a median value ($a \rightarrow \infty$ would imply determining $\rho$ purely as the average leading to zero bias for the sample without any hard-scatter event).

\section{Performance using the framework from the 2014 Pileup Workshop}
\label{app:framework_pileup_workshop}
We compare the performance of the new method with PUPPI method \cite{Bertolini:2014bba} using the common open-source software framework~\cite{PileupWorkshopFramework} defined at the ``Pileup Workshop'' held in May 2014 at CERN~\cite{PileupWorkshopIndico}. The code used to obtain the results in this section is located in the folder \texttt{comparisons/SignalFreeBackgroundEstimator} of this framework, version 1.1.1. The purpose of using this framework is to take advantage of the software implementation of PUPPI which is publicly only available within this framework. The main difference between PUPPI and CS-based methods is that PUPPI does not use $\rho$. Therefore, PUPPI can be used to:
\begin{itemize}
 \item compare the new method with a method independent on $\rho$;
 \item investigate the performance of the new method in the case that the seeds in Step 1 are obtained with a method independent on $\rho$.
\end{itemize}

The samples used for the comparison presented here are available from \bibref{PileupWorkshopSamples}. Four hard-scatter physics processes overlaid with a fixed number of \pileup events are used. The four hard-scatter samples are dijet events with at least one reconstructed \aktfour jet satisfying $\pt\geq$ 20, 50, 100, or 500\gev. $B$-hadrons are kept stable, and UE is not simulated. The hard-scatter and \pileup samples are simulated using Pythia 8.185 with tune 4C using proton-proton collisions at $\sqrt{s}=14\tev$. Four \pileup conditions are used: $\npu=$ 30, 60, 100, and 140. The performance is evaluated for \aktfour jets.
The mass of all particles is set to zero, preserving \pt, rapidity, and azimuth. Only particles with $|y|<4$ are used. The corrections are done on events without any detector simulation but assuming an idealized tracker, where the information about the origin of charged particles (\pileup or hard-scatter event) is known. In that way, the charged \pileup particles can be directly removed, and/or the information about charged particles can be further used to correct the neutral particles. 

We compare the ICS algorithm employing three different methods to estimate $\rho$ and PUPPI. The configuration of these four algorithms is following:
\begin{itemize}
  \item \textbf{ICS grid-median:} All charged \pileup particles are discarded. Only the neutral particles are corrected using the Iterative Constituent Subtraction method with two iterations and without ghost removal. Parameters: $\alpha=1$ and $\Aghost=0.005$ for both iterations, $\DeltaRmax=0.2$ and $\DeltaRmax=0.12$ for the first and the second iteration, respectively. To estimate $\rho$, the standard grid-median approach with the recommended setting is used (grid size of 0.55).
  \item \textbf{ICS signal-free (seeds from ICS)} - Same configuration as in the previous method, but the new method is used to estimate $\rho$ (instead of the grid-median method). The used parameters are: $a=1.2$, $W=0.8$, floating $C$ with $K=0.1$ and $\Smin=0.1$. The signal seeds for Step 1 were obtained using a pile-corrected event with the ICS grid-median method.
  \item \textbf{ICS signal-free (seeds from PUPPI)} - Same configuration as in the previous method, but the signal seeds for Step 1 are obtained using pileup-corrected event with PUPPI (instead of ICS).
  \item \textbf{PUPPI:} The implementation is taken from the \texttt{comparisons/review} in \bibref{PileupWorkshopFramework} version 1.0.0, which is the original implementation provided by the PUPPI authors in the context of the 2014 Pileup Workshop.
\end{itemize}

To compare the performance of the methods, the average bias and resolution for the jet \pt are evaluated. From each event, all jets with $\pt\in[20,50]\gev$ and $|y|<2.5$ are selected. Each corrected jet is matched to the closest true jet by requiring $\sqrt{\Delta y^2+\Delta\phi^2}<0.3$ (the matching efficiency is above $99.5\%$). To follow recommendations from the workshop, the average jet \pt bias, $\langle\Delta\pt\rangle$, and \pt resolution, $\sigma_{\Delta\pt}$, is evaluated as the average and RMS from the \pt difference between true and matched corrected jet \pt, respectively. 

\begin{figure}[!h]
  \centering
    \includegraphics[angle=270,width=\textwidth]{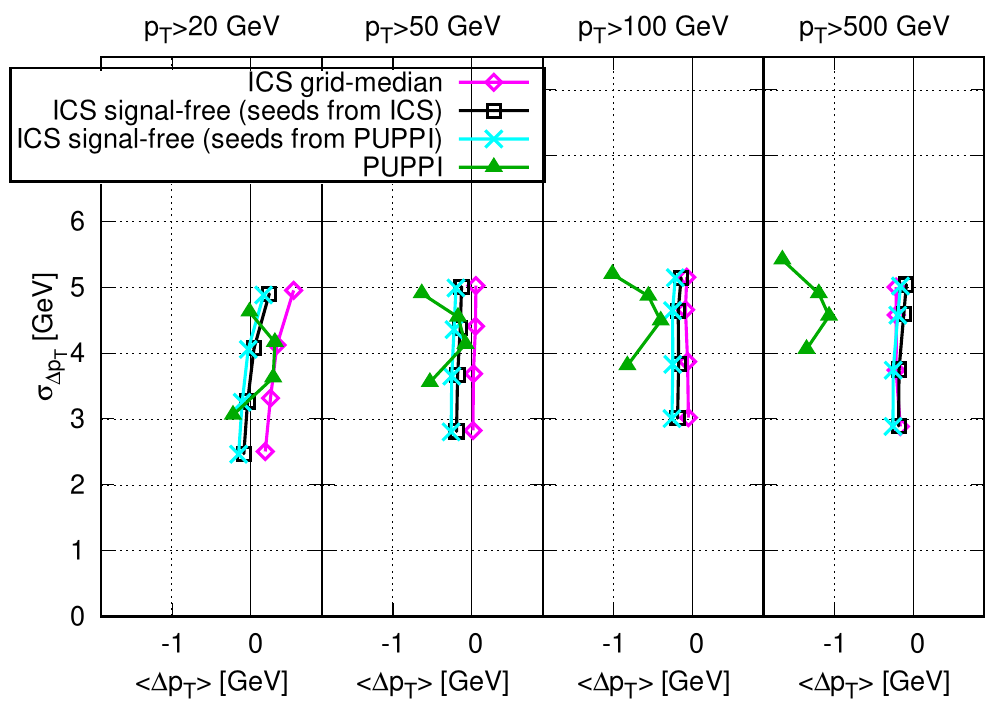}
  \caption{Comparison of the jet \pt resolution as a function of jet \pt bias for PUPPI and the ICS method with three different $\rho$ estimation methods: grid-median background estimation, signal-free background estimation using seeds from ICS, and signal-free background estimation using seeds from PUPPI. Dijet events are used with different jet \pt cuts in each panel. Each curve corresponds to a different method, and the 4 points on each curve correspond to $\npu$ = 30, 60, 100, and 140 from bottom to top.}
  \label{fig:2014PileupWorkshop}
\end{figure}

The comparison of the above methods is shown in \figref{2014PileupWorkshop}. One can observe the following:
\begin{itemize}
\item The performance of the two ICS methods with signal-free background estimator, which uses different strategies to get signal seeds, is very similar. The use of PUPPI to deliver seeds does not bring an improvement.

\item Qualitatively, one can see a similar improvement when using the new method over the grid-median method as discussed in \secref{jet}. The jet \pt bias for the ICS with the signal-free method remains roughly the same for all four samples. On the contrary, the bias changes by about 0.5 GeV for the ICS grid-median method when comparing the four samples for pileup conditions with $\npu=140$. The quantitative difference between grid-median and signal-free methods is not so large as in \secref{jet} due to much lower jet multiplicity differences in the four hard-scatter samples used here. The average jet multiplicity in the dijet samples used here increases from 1.6 to 4.4 for jets with $\pt>20$ and $\pt>500$\gev, respectively.

\item Despite the fact that PUPPI does not use $\rho$, the jet \pt bias obtained with PUPPI changes by 1.5 GeV for pileup conditions with $\npu=140$ when comparing the four samples. Based on these studies, one can conclude that PUPPI has a significantly larger dependence on the hard-scatter event than the ICS signal-free method.
\end{itemize}

%This is the most common positions for acknowledgments. A macro is
%available to maintain the same layout and spelling of the heading.

%\paragraph{Note added.} This is also a good position for notes added
%after the paper has been written.


\bibliography{ms}
\bibliographystyle{unsrt}

% \begin{thebibliography}{99}
%
% examples:
%
%\bibitem{a}
%Author, \emph{Title}, \emph{J. Abbrev.} {\bf vol} (year) pg.
%
%\bibitem{b}
%Author, \emph{Title},
%arxiv:1234.5678.
%
%\bibitem{c}
%Author, \emph{Title},
%Publisher (year).

% Please avoid comments such as "For a review'', "For some examples",
% "and references therein" or move them in the text. In general,
% please leave only references in the bibliography and move all
% accessory text in footnotes.

% Also, please have only one work for each \bibitem.

%\end{thebibliography}

\end{document}